\documentclass[prd, preprint,
superscriptaddress,tightenlines,nofootinbib, eqsecnum]{revtex4-2}

\usepackage{amsmath}
\usepackage{amsfonts}
\usepackage{amssymb}
\usepackage{bm}
\usepackage{hyperref}
\usepackage{subcaption}
\usepackage{mathrsfs}
\usepackage{graphicx}
\usepackage{diagbox}
\usepackage{empheq}
\usepackage{multirow}
\usepackage{float}
\usepackage{graphicx}
\usepackage{ulem}
\usepackage[usenames]{color}
\usepackage[compat=1.0.0]{tikz-feynman}

\allowdisplaybreaks

\hypersetup{
 bookmarks=true,         
 unicode=false,          
 pdftoolbar=true,        
 pdfmenubar=true,        
 pdffitwindow=false,     
 pdfstartview={FitH},    
 pdftitle={My title},    
 pdfauthor={Author},     
 pdfsubject={Subject},   
 pdfcreator={Creator},   
 pdfproducer={Producer}, 
 pdfkeywords={keyword1} {key2} {key3}, 
 pdfnewwindow=true,      
 colorlinks=true,       
 linkcolor=red,          
 citecolor=cyan,        
 filecolor=magenta,      
 urlcolor=darkgreen,           
 linktocpage=true
}

\definecolor{darkgreen}{rgb}{0,0.5,0}

\newcommand{\FPprop}{\mathop{\mathrm{FP}}_{B=0}\Box^{-1}_\text{ret}}

\def\pf{\underset{B=0}{\mathrm{FP}}}

\newcommand{\be}{\begin{equation}}
	\newcommand{\ee}{\end{equation}}
\newcommand{\sbe}{\begin{subequations}}
	\newcommand{\see}{\end{subequations}}

\newcommand{\ba}{\begin{eqnarray}}
	\newcommand{\ea}{\end{eqnarray}}
\newcommand{\p}{\partial}

\newcommand{\nn}{\nonumber}

\DeclareSymbolFontAlphabet{\mathrsfs}{rsfs}
\DeclareMathAlphabet{\mathcal}{OMS}{cmsy}{m}{n}

\newcommand\calO{{\mathcal{O}}}

\newcommand{\dd}{\mathrm{d}}

\newcommand{\dM}{\mathrm{M}}
\newcommand{\dMbar}{\overline{\mathrm{M}}}
\newcommand{\dS}{\mathrm{S}}
\newcommand{\dSbar}{\overline{\mathrm{S}}}

\allowdisplaybreaks

\DeclareSymbolFontAlphabet{\mathrsfs}{rsfs}
\DeclareMathAlphabet{\mathcal}{OMS}{cmsy}{m}{n}

\begin{document}
	
\title{Gravitational-wave tails of memory}

\author{David \textsc{Trestini}}\email{david.trestini@obspm.fr}
\affiliation{Laboratoire Univers et Théories, Observatoire de Paris, Université PSL, Université Paris Cité, CNRS, F-92190 Meudon, France}
\affiliation{$\mathcal{G}\mathbb{R}\varepsilon{\mathbb{C}}\mathcal{O}$, 
	Institut d'Astrophysique de Paris,\\ UMR 7095, CNRS, Sorbonne Universit{\'e},\\
	98\textsuperscript{bis} boulevard Arago, 75014 Paris, France}

\author{Luc \textsc{Blanchet}}\email{luc.blanchet@iap.fr}
\affiliation{$\mathcal{G}\mathbb{R}\varepsilon{\mathbb{C}}\mathcal{O}$, 
	Institut d'Astrophysique de Paris,\\ UMR 7095, CNRS, Sorbonne Universit{\'e},\\
	98\textsuperscript{bis} boulevard Arago, 75014 Paris, France}

\date{\today}

\begin{abstract}
Gravitational-wave tails are linear waves that backscatter on the curvature of space-time generated by the total mass-energy of the source. The non-linear memory effect arises from gravitational waves sourced by the stress-energy distribution of linear waves themselves. These two effects are due to quadratic multipolar interactions (mass-quadrupole and quadrupole-quadrupole) and are well known. Also known are the tails generated by tails themselves (cubic ``tails-of-tails'') and the tails generated by tails-of-tails or \emph{vice versa} (quartic ``tails-of-tails-of-tails''). In this work, we focus on the cubic ``tails-of-memory'' corresponding to the mass-quadrupole-quadrupole interaction, as well as the ``spin-quadrupole tails'', which are due to the cubic interaction between the mass, the total angular momentum and the quadrupole. The tails-of-memory and the spin-quadrupole tails contribute to the asymptotic waveform at the fourth-post-Newtonian (4PN) order beyond quadrupolar radiation.
\end{abstract}


\maketitle

\section{Introduction}
\label{sec:intro}


Due to the non-linear nature of general relativity, non-linear effects in the propagation of gravitational waves from their source to a distant detector play an important role in predictions for gravitational waves generated by compact binary systems~\cite{Maggiore,BuonSathya15,BlanchetLR,GWroadmap}. The prototype of such effects is the gravitational-wave \emph{tail}, namely the backscattering of linear waves on the curvature of space-time generated by the total mass-energy of the source~\cite{Bo59,ThK75,Th80,BD88}. The tail effect has actually already  been detected by the LIGO-Virgo observations~\cite{LIGOtestGR}. This constitutes an interesting test of the non-linear structure of general relativity~\cite{BSat94,BSat95,AIQS06a,AIQS06b}.

At leading order the tail effect is a quadratic coupling between the Arnowitt-Deser-Misner~(ADM) total mass of space-time and the mass quadrupole moment of the source. This effect appears at one-and-a-half post-Newtonian (1.5PN) order in the waveform beyond the Einstein quadrupole formula~\cite{BD92,P93a,Wi93,BS93,FStail,GLPR16}. At 3PN order there is the \emph{tail-of-tail} effect which is a cubic coupling between two ADM masses and the quadrupole~\cite{B98tail,FBI15}. The previous effects have been included into the PN templates for compact binary inspiral that are routinely used by the LIGO-Virgo detectors. Furthermore, at quartic order there is the \emph{tail-of-tail-of-tail} which arises at 4.5PN order and involves three masses and a quadrupole~\cite{MBF16,MNagar17}. 

The non-linear memory effect, \emph{i.e.} the permanent change in the wave amplitude after the passage of a gravitational wave burst, is due to the re-radiation of quadrupole gravitational waves by a linear quadrupole wave and corresponds to the coupling between two quadrupole moments~\cite{B90,Chr91,WW91,Th92,BD92,B98quad,F09,F11,N17spinmemory}. It enters at 2.5PN order in the waveform. The oscillatory (``AC'') piece of this interaction is routinely used by the LIGO-Virgo detector, but the genuine, secular memory effect (which is almost zero frequency or ``DC'') has not yet been detected experimentally, but could be observed in the coming years with ground-based gravitational wave detectors~\cite{Favata:2009ii,Lasky:2016knh,McNeill:2017uvq}.

\begin{figure}[H]\centering
\begin{subfigure}[b]{0.25\textwidth}
\includegraphics[width=\linewidth]{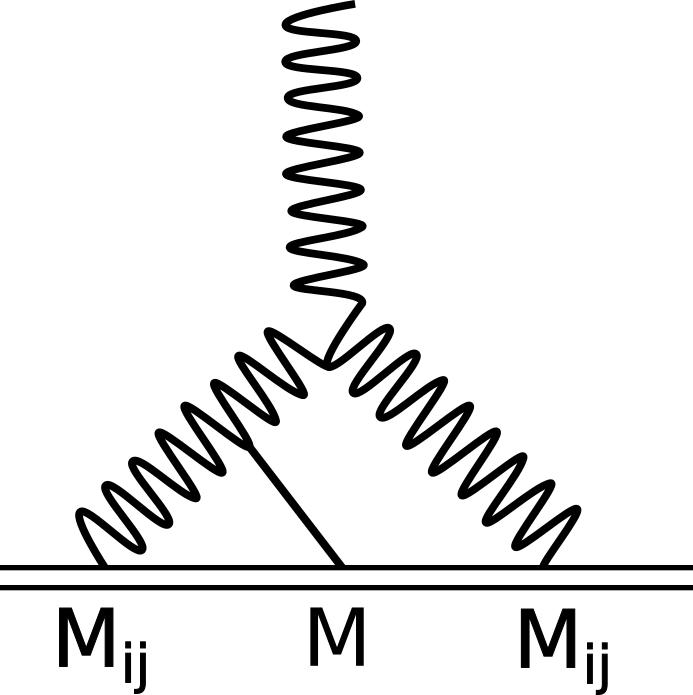}\caption{}\label{subfig:Feynman1}
\end{subfigure}
\qquad\qquad
\begin{subfigure}[b]{0.25\textwidth}
\includegraphics[width=\linewidth]{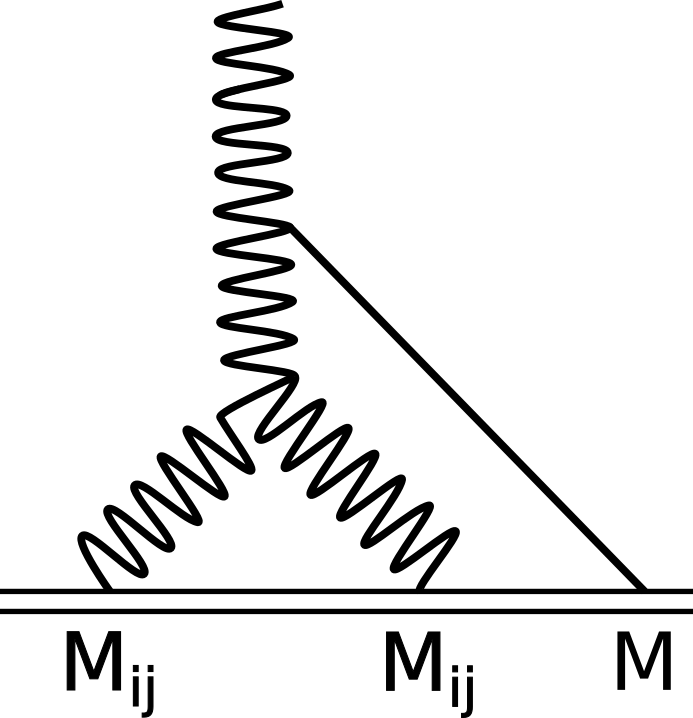}\caption{}\label{subfig:Feynman2}
\end{subfigure}
\qquad\qquad
\begin{subfigure}[b]{0.25\textwidth}
\includegraphics[width=\linewidth]{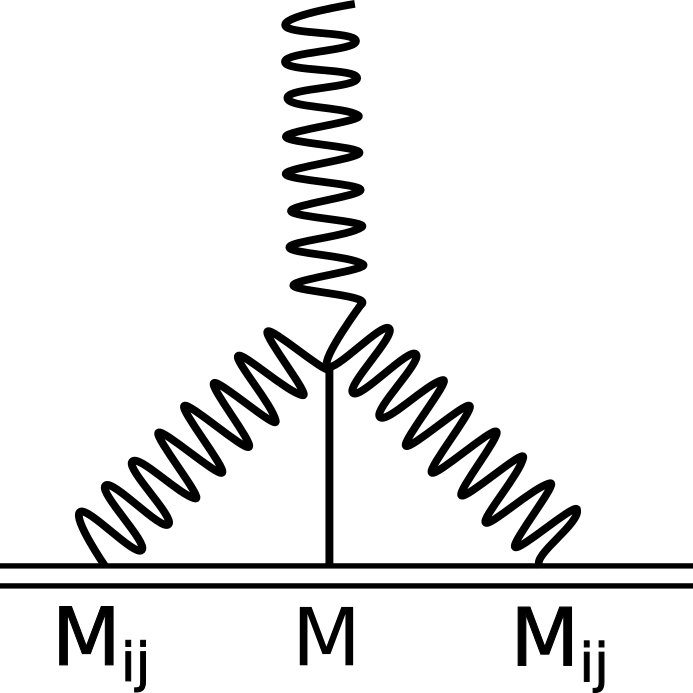}\caption{}\label{subfig:Feynman3}
\end{subfigure}
\caption{Feynman diagrams associated to the tails-of-memory.}
\label{fig:Feynman}
\end{figure}

In the present paper we shall proceed one step further and compute the cubic non-linear interaction between the constant ADM mass $\dM$ and two quadrupole moments $\dM_{ij}$, which we denote \mbox{as $\dM\times\dM_{ij}\times \dM_{ij}$}. Physically this can be viewed as a combination between the tails produced by the memory and the memory associated with the tail; we coin this effect ``\emph{tails-of-memory}''. From an effective field theory (EFT) perspective, the tails-of-memory correspond to the three Feynman diagrams shown in Fig.~\ref{fig:Feynman}, which we consider here for illustrative purposes; see~\cite{GR06,FSrevue,Porto16,Levi18} for their precise computational meaning within the EFT framework.

Interestingly, the tails-of-memory arise at 4PN order in the waveform, \emph{i.e.}, at the level which is our overall goal currently~\cite{MHLMFB20,HFB_courant,MQ4PN_IR,MQ4PN_renorm,MQ4PN_jauge,TLB22}, motivated by third generation detectors on the ground and by LISA in space. For this reason, we also compute the only other cubic contribution that arises at 4PN order, namely the ``\emph{spin-quadrupole tail}''  interaction $\dM\times\dS_i\times \dM_{ij}$ between the constant ADM mass, the constant ADM angular momentum and a quadrupole moment. 

Note that the contributions due to the dimensional regularization of all non-linear interactions relevant at 4PN order (in particular the tails-of-memory) have already been included into the definition of a ``renormalized'' source quadrupole moment, see~\cite{MQ4PN_renorm} for details. Therefore we shall compute the tails-of-memory and spin-quadrupole tails purely in three spatial dimensions; adding them to the renormalized source quadrupole moment~\cite{MQ4PN_renorm} will naturally yield the physical radiative moment which is observed at infinity.

Since the tails-of-memory arise at 4PN order in the waveform, and that radiation reaction adds another 2.5PN order, we expect the tails-of-memory to contribute in the form of radiation modes to the conservative equations of motion and Lagrangian/Hamiltonian at 6.5PN order. This is far beyond the state-of-the-art on equations of motion which is 4PN order~\cite{DJS14,BBBFMa,MBBF17,FS19,FPRS19,Blumlein20} (see also the results~\cite{Foffa5PN,bini2020binary,Blumlein21,almeida2021tail} at 5PN order and~\cite{bini2020sixth} at 6PN order).

We shall compute the tails-of-memory using the multipolar-post-Minkowskian (MPM) formalism~\cite{BD86,B87} but with an important variant with respect to previous works such as~\cite{B98quad,B98tail}: we shall construct the metric directly in \emph{radiative} coordinates (see~\cite{Papa69, MadoreI, MadoreII} for a general definition), following the method proposed in~\cite{B87}, instead of the usual construction in harmonic coordinates. The great advantage is that we shall avoid the appearance of the far zone logarithms that plague harmonic coordinates. A minor disadvantage is that we shall have to apply a correction to account for the different expressions of the multipole moments in radiative and harmonic coordinates, but this correction is already known~\cite{TLB22}.

Finally, despite the fact that we are mainly motivated by the case of quasi-circular compact binaries, the results presented in this paper are valid for a generic isolated source described in terms of its multipole moments; the source could be a general hydrodynamical fluid with compact support, and it needs not even be a post-Newtonian source.

The plan of this paper is as follows. In Sec.~\ref{sec:radMPM}, we review the radiative version of the MPM algorithm. In Sec.~\ref{sec:quadratic}, we present our results for the radiative metric at quadratic order, and discuss the structure of the cubic source. In Sec~\ref{sec:techniques}, we present a novel integration technique for the cubic source, and apply it as a check to the known cases of tails-of-tails and tails. Then, in Sec.~\ref{sec:radiativeQuadrupole}, we present our ``raw'' result in the form of complicated functionals of the multipole moments, along with a set of coefficients given by Table~\ref{table:coeffs}. We then introduce a method of simplification, which drastically reduces the ``raw'' result, and allows for a test of our integration method. Finally, we present in Sec.~\ref{sec:results} the final cubic radiative quadrupole moment (in its simplified form) in terms of the harmonic canonical moments. Appendix~\ref{app:NPQR} presents a practical method for extracting multipole moments, and  Appendix~\ref{app:convergence} presents a proof for the convergence of our final integrals.

\section{The radiative MPM algorithm}
\label{sec:radMPM}

The vacuum field equations of general relativity are recast in the form of a wave like equation by defining the gothic metric deviation $h^{\mu\nu} \equiv \sqrt{-g} g^{\mu\nu} - \eta^{\mu\nu}$, where $\eta^{\mu\nu}$ is the flat Minkowski metric (in the mostly plus signature), $g^{\mu\nu}$ is the inverse of the usual covariant metric and $g \equiv \det(g_{\mu\nu})$ is the determinant. In a general coordinate system, the vacuum field equations become
\begin{equation}\label{eq:EinsteinEquations} 
	\Box h^{\mu\nu} - \partial H^{\mu\nu} = \Lambda^{\mu\nu}\bigl[h, \partial h, \partial^2 h\bigr]\,,\end{equation}
where $\Box$ is the flat d'Alembertian operator, the divergence of the metric is $H^{\mu} \equiv \partial_\nu h^{\mu\nu}$ and we introduce the useful shorthand notation 
$\partial \theta^{\mu\nu} \equiv 2\partial^{(\mu} \theta^{\nu)}-\eta^{\mu\nu}\partial_\rho \theta^\rho$, valid for any vector~$\theta^\mu$. The non-linear source term in vacuum reads explicitly 
\begin{equation}
\label{eq:LambdaRad}
\Lambda^{\mu\nu} =  \Lambda_{\text{harm}}^{\mu\nu} + 2 h^{\rho(\mu}\partial_\rho H^{\nu)}- \partial_\rho \left(h^{\mu\nu} H^\rho \right)\,,
\end{equation}
where  $\Lambda_{\text{harm}}^{\mu\nu}$ is the source term in a harmonic gauge satisfying $H^\mu=0$ and is given explicitly by Eq.~(24) of~\cite{BlanchetLR}. Taking the divergence of~\eqref{eq:EinsteinEquations}, and using the fact that $\partial_\nu \partial H^{\mu\nu}=\Box H^\mu$, we find that the source term is identically divergenceless in any coordinate system, \emph{i.e.}, $\partial_\nu \Lambda^{\mu\nu} = 0$, which is equivalent to the contracted Bianchi identity.

From now on, $h^{\mu\nu}$ will denote the metric in a radiative type coordinate system~\cite{Papa69, MadoreI, MadoreII}, characterized by the absence of logarithms of the radial distance $r$ in the far retarded zone, \emph{i.e.}, when $r\to+\infty$ with $t-r=$ const. With the MPM construction the radiative metric is formally decomposed to any post-Minkowskian order as 
\begin{equation}\label{eq:hMPM} 
	h ^{\mu\nu} = \sum_{n=1}^{+\infty} G^n h_n^{\mu\nu} \,, 
\end{equation}
where each of the coefficients will be a functional of two sets of STF multipole moments $\dMbar_L$ and $\dSbar_L$ which by definition parametrize the linearized approximation $h_1^{\mu\nu}$ defined as
\begin{equation}\label{eq:h1}
	h^{\mu\nu}_1 \equiv h_{\text{can}\,1}^{\mu\nu}[\dMbar_L, \dSbar_L] + \partial \xi_1^{\mu\nu}\,, 
\end{equation}
where the ``canonical'' linear metric reads explicitly~\cite{SB58, Pi64, Th80, BD86}\footnote{We use the symmetric-trace-free (STF) formalism, with $L=i_1\cdots i_\ell$ a multi-index with $\ell$ spatial indices, and $\hat{n}_L=\text{STF}[n_{i_1}\cdots n_{i_\ell}]$ the usual STF harmonics of order $\ell$. Similarly we shall denote $\hat{\partial}_L=\text{STF}[\partial_{i_1}\cdots \partial_{i_\ell}]$. We generally pose $c=1$.} 
\newpage
\begin{subequations}\label{eq:h1can}
	\begin{align}
		h_{\text{can}\,1}^{00}[\dMbar_L, \dSbar_L] &= - 4\sum_{\ell \geqslant 0}\frac{(-)^\ell}{\ell !}\partial_L\left[\frac{1}{r}\,\dMbar_L\left(t-r\right)\right]\,,\\
		h_{\text{can}\,1}^{0i} [\dMbar_L, \dSbar_L]&=4\sum_{\ell \geqslant 1}\frac{(-)^\ell}{\ell !}\left\lbrace\partial_{L-1}\left[\frac{1}{r}\,\dMbar_{iL-1}^{(1)}\left(t-r\right)\right]+ \frac{\ell}{\ell+1}\epsilon_{iab} \partial_{aL-1}\left[\frac{1}{r}\,\dSbar_{b L-1}\left(t-r\right)\right]\right\rbrace\,,\\
		h_{\text{can}\,1}^{ij}[\dMbar_L, \dSbar_L] &= -4\sum_{\ell \geqslant 2}\frac{(-)^\ell}{\ell !}\left\lbrace\partial_{L-2}\left[\frac{1}{r}\,\dMbar_{ijL-2}^{(2)}\left(t-r\right)\right]+ \frac{2\ell}{\ell+1}\partial_{aL-2}\left[\frac{1}{r}\,\epsilon_{ab(i}\dSbar_{j) b L-2 }^{(1)}\left(t-r\right)\right]\right\rbrace\,.
	\end{align}
\end{subequations}
It satisfies the harmonic gauge condition, $\partial_\nu h_{\text{can}\,1}^{\mu\nu}=0$. The linear gauge transformation vector in~\eqref{eq:h1} is
\begin{equation}\label{eq:xi1}
	\xi_1^{\mu} = 2M \eta^{0\mu} \ln\left(\frac{r}{b_0}\right)\,,
\end{equation}
with $b_0$ an arbitrary constant length scale. The role of this gauge transformation is to ensure that the retarded time $u=t-r$ is asymptotically a null coordinate when $r\to+\infty$. 

The multipole moments $\dMbar_L$ and $\dSbar_L$ will depend specifically of the radiative MPM construction that we shall perform. In particular, as we have shown in~\cite{TLB22}, the moments $\dMbar_L$ and $\dSbar_L$ will differ from the set of multipole moments $\dM_L$ and $\dS_L$ defined in the MPM construction in harmonic coordinates, so we shall have to correct for that difference using the results of~\cite{TLB22}. The moments in harmonic coordinates $\dM_L$ and $\dS_L$ can be directly related the to source \emph{via} asymptotic matching, and are known to 4PN order~\cite{MHLMFB20, MQ4PN_IR, MQ4PN_renorm, MQ4PN_jauge}. Consistently the relation between the moments $\{\dMbar_L, \dSbar_L\}$ and the harmonic coordinate moments $\{\dM_L, \dS_L\}$ has been investigated, and in the case of the quadrupole moment, determined to 4PN order in~\cite{TLB22}; we recall the result in Eq.~\eqref{eq:momentRedefinition} below. Note also that the conserved quantities are equal in both coordinate systems, \emph{i.e.}, $\dM = \dMbar$ and $\dS_i = \dSbar_i$, so in these cases we will use both notations interchangeably.

Once the linearized solution is defined, the non-linear corrections are readily obtained by injecting the PM expansion into the field equations and solving these equations iteratively over the PM order. At any order $n$, the general equation to solve is
\begin{equation}\label{eq:EinsteinEquationOrderN} 
	\Box h_n^{\mu\nu} - \partial H^{\mu\nu}_{n} = \Lambda_n^{\mu\nu}\,,
\end{equation}
where $\Lambda_n^{\mu\nu}\equiv\Lambda^{\mu\nu}[h_1, \cdots, h_{n-1}]$ is built out from previous iterations. We first construct a particular retarded solution to the wave equation, satisfying $\Box u_n^{\mu\nu} = \Lambda_n^{\mu\nu}$, as
\begin{equation}\label{eq:unMPM}
	u_n^{\mu\nu} = \FPprop \biggl[\left(\frac{r}{r_0}\right)^{B} \Lambda_{n}^{\mu\nu}\biggr]\,,\end{equation}
where $\Box^{-1}_\text{ret}$ is the standard retarded integral operator, and a particular regularization with regulator $(r/r_0)^B$, where $B$ is complex and $r_0$ a constant length scale, is introduced to cope with the singularity of the multipole expansion when $r\to 0$. The finite part (FP) when $B\to 0$, \emph{i.e.}, the coefficient of the zeroth power of $B$ in the Laurent expansion when $B\to 0$, provides the requested solution in this context.

Next we compute its divergence $w_n^\mu \equiv \partial_\nu u_n^{\mu\nu}$, and from that divergence we are able to construct an homogeneous solution $v_n^{\mu\nu}$ to the wave solution, which satisfies at once $\Box v_n^{\mu\nu}=0$ and $\partial_\nu v_n^{\mu\nu}=- w_n^\mu$. We describe in Appendix~\ref{app:NPQR} the practical method which permits to obtain $v_n^{\mu\nu}$ starting from $w_n^{\mu}$. At this point it is clear that the sum $s_n^{\mu\nu}=u_n^{\mu\nu}+v_n^{\mu\nu}$ satisfies the field equation~\eqref{eq:EinsteinEquationOrderN}, since we have $\Box s_n^{\mu\nu} = \Lambda_n^{\mu\nu}$ and $\partial_\nu s_n^{\mu\nu} = 0$. However in order to obtain the sought-for radiative metric, we still have to correct for the light cone deviation to order $n$, and for that we apply a linear gauge transformation parametrized by a vector $\xi_n^\mu$, hence defining
\begin{equation}\label{eq:hnmunu}
	h_n^{\mu\nu} \equiv u_n^{\mu\nu}+v_n^{\mu\nu} +\partial \xi_n^{\mu\nu}\,.
\end{equation}

The construction of this gauge vector crucially relies on the particular structure of the source term, proved by induction over the order $n$, \emph{i.e.}, an asymptotic expansion when $r\to+\infty$ with $t-r=$ const in simple powers of $1/r$ (without logarithms) whose leading piece $1/r^2$ takes the particular form
\begin{equation}\label{eq:Lambdanmunu}
	\Lambda_n^{\mu\nu} = \frac{k^\mu k^\nu}{r^2} \sigma_n(u,\mathbf{n}) + \calO\left(\frac{1}{r^3}\right) \,,
\end{equation}
where $k^\mu = (1,\mathbf{n})$ is the outgoing Minkowskian vector. An appropriate choice (but by no means unique) for the gauge vector is 
\begin{equation}\label{eq:xinmu}
	\xi_n^{\mu\nu} = \FPprop \biggl[ \left(\frac{r}{r_0}\right)^{B} \frac{k^\mu}{2 r^2} \int_0^{+\infty} \dd \tau\, \sigma_n(u-\tau, \mathbf{n})\biggr]\,.
\end{equation}
With this choice of gauge vector, one can prove by induction that the metric is free of far-zone logarithms at all post-Minkowskian orders~\cite{B87, TLB22}, and thus falls into the class of radiative coordinate systems for which the extraction of physical observables at infinity is easy. The implementation of the above construction for the case of the tails-of-memory will indeed prove very important. 

\section{Results at quadratic order}
\label{sec:quadratic}

In the present paper we shall compute from first principles the following three \emph{cubic} non-linear multipole interactions (\textit{i.e.}, $n=3$ in the PM expansion): 
\begin{equation*}
	\dM^2 \times \dMbar_{ij}\,,\qquad \d M \times \dS_i \times \dMbar_{ij}\quad\text{and}\quad  \d M \times \dMbar_{ij} \times \dMbar_{ij}\,.
\end{equation*}
The first one is known as the quadrupole ``tail-of-tail''~\cite{B98tail}, the second one can be called the ``spin-quadrupole tail'' while the third one is the quadrupole ``tail-of-memory'' interaction which is our main goal here. To reach these goals we evidently need to control the following quadratic interactions:
\begin{equation*}
	\dM\times \dM \,,\qquad \dM\times \dS_i \,, \qquad \dM\times\dMbar_{ij}\,, \qquad \dMbar_{ij}\times \dMbar_{ij} \quad  \text{and}\quad \dS_i \times \dMbar_{ij} \,,
\end{equation*}
which we shall compute at any distance $r$ from the source, while the cubic interactions will just be computed asymptotically when $r\to+\infty$. 

\subsection{Linear and quadratic results}
\label{subsec:quadratic}

The linear interactions $\dM$, $\dS_i$ and $\dMbar_{ij}$ are read off directly from the linear metric~\eqref{eq:h1}--\eqref{eq:h1can}, including the gauge transformation~\eqref{eq:xi1}:
\newpage
\begin{subequations}
	\label{eq:hmunuLinear}
	\begin{align}
		\label{subeq:hmunuM}
		h_\dM^{00} &= - \frac{4\dM}{r}\,, & h_\dM^{0i} &=  - \frac{2 \dM n^i}{r}\,, & h_\dM^{ij} &= 0\,,\\
		\label{subeq:hmunuSi}
		h_{\dS_{i}}^{00} &= 0\,, & h_{\dS_{i}}^{0i} &= \frac{2 n^a \dS_{i\vert a}}{r^2}\,, & h_{\dS_{i}}^{ij} &= 0\,,\\
		\label{subeq:hmunuMbarij}
		h_{\dMbar_{ij}}^{00} &= - 2 \partial_{ab} \biggl[\frac{\dMbar_ {ij}}{r}\biggr]\,,  &h_{\dMbar_{ij}}^{0i} &=   2 \partial_{a} \biggl[\frac{\dMbar_{ia}^{(1)}}{r}\biggr]\,,  & h_{\dMbar_{ij}}^{ij} &= - \frac{2 \dMbar_{ij}^{(2)}}{r}\,,
	\end{align}
\end{subequations}
where we have henceforth posed $\dS_{i\vert j} \equiv \epsilon_{ija} \dS_a$. 

In the harmonic construction, the quadratic metrics that are known are $\dM\times\dM$~\cite{B98tail}, $\dM\times\dM_{ij}$~\cite{BD92} and $\dM_{ij}\times \dM_{ij}$~\cite{B98quad}. We also know the $ \dS_i \times \dM_{ij}$ metric asymptotically, to leading order in~$1/r$~\cite{B98quad}. 

The radiative metrics for the static $\dM\times \dM$ and stationary $\dM \times \dS_i $ interactions, that are respectively part of the Schwarzschild and Kerr metrics (in ``radiative'' coordinates), \mbox{read \cite{SuppMaterial}}
\begin{subequations}
\begin{align}
	\label{subeq:hmunuMxM}
	h_{\dM\times\dM}^{00} &= - \frac{3\dM^2}{r^{2}}\,,&
	h_{\dM\times\dM}^{0i} &= 0 \,,&
	h_{\dM\times\dM}^{ij} &= -\frac{\dM^2 n^{ij}}{r^{2}}\,,\\
	\label{subeq:hmunuMxSi}
	h_{\dM\times\dS_i}^{00} &= 0\,,&
	h_{\dM\times\dS_i}^{0i} &= \frac{2\dM}{r^{3}} n_a \dS_{i\vert a} \,,&
	h_{\dM\times\dS_i}^{ij} &= 0\,.
\end{align}
\end{subequations}
Note that although $\dMbar = \dM$ as we said, the $\dM\times\dM$ radiative metric differs from the harmonic one, see Eqs.~(2.7) in~\cite{B98tail}. In the radiative construction, the $\dM\times \dMbar_{ij}$ interaction is given in~\cite{TLB22} and we reproduce it here for completeness \cite{SuppMaterial}:
\begin{subequations}\label{eq:hmunuMxMbarij}
	\begin{align}
	\label{subeq:h00MxMbarij}
		h_{\dM\times \dMbar_{ij}}^{00} &= 8 \dM  n_{ab}   \int_1^{+\infty} \dd x \,\overline{Q}_2(x,r)\,\dMbar^{(4)}_{ab}(t-rx)\nonumber\\
		&\quad+ \frac{\dM n_{ab}}{r}\Big(\frac{118}{15}  \dMbar^{(3)}_{ab} + \frac{23}{5} r^{-1} \dMbar^{(2)}_{ab}- \frac{117}{5} r^{-2}  \dMbar^{(1)}_{ab} -21 r^{-3}  \dMbar_{ab}\Big)\,,\\
	\label{subeq:h0iMxMbarij}
		h_{\dM\times \dMbar_{ij}}^{0i} &= 8 \dM  n_{a}   \int_1^{+\infty} \dd x \,\overline{Q}_1(x,r)\, \dMbar^{(4)}_{ai}(t-rx)\nonumber\\
		&\quad +\frac{\dM n_{iab}}{r}\bigg[\dMbar^{(3)}_{ab}+3 r^{-1}\dMbar^{(2)}_{ab}- r^{-2}\dMbar^{(1)}_{ab} \bigg]\nonumber\\
		&\quad+ \frac{\dM n_{a}}{r}\bigg[\frac{43}{15}\dMbar^{(3)}_{ai}-\frac{107}{15} r^{-1}\dMbar^{(2)}_{ai}- 5r^{-2}\dMbar^{(1)}_{ai} \bigg]\,,\\
	\label{subeq:hijMxMbarij}
		h_{\dM\times \dMbar_{ij}}^{ij} &= 8 \dM \int_1^{+\infty} \dd x \,\overline{Q}_0(x,r)\, \dMbar^{(4)}_{ij}(t-rx) \nonumber\\
		&\quad +\frac{\dM n_{ijab}}{r}\bigg[ -\frac{1}{2}\dMbar^{(3)}_{ab}-3r^{-1}\dMbar^{(2)}_{ab}-\frac{15}{2} r^{-2}\dMbar^{(1)}_{ab}-\frac{15}{2} r^{-3}\dMbar_{ab}\bigg]\nonumber\\
		&\quad +\frac{\dM \delta_{ij} n_{ab}}{r}\bigg[ -\frac{1}{2}\dMbar^{(3)}_{ab}-2r^{-1}\dMbar^{(2)}_{ab}-\frac{1}{2} r^{-2}\dMbar^{(1)}_{ab}-\frac{1}{2} r^{-3}\dMbar_{ab}\bigg]
		\nonumber\\
		&\quad +\frac{\dM n_{a(i}}{r}\bigg[4\dMbar^{(3)}_{j)a}+6 r^{-1}\dMbar^{(2)}_{j)a}+6 r^{-2}\dMbar^{(1)}_{j)a}+6 r^{-3}\dMbar_{j)a}\bigg]\nonumber\\
		&\quad +\frac{\dM }{r}\bigg[ -\frac{107}{15}\dMbar^{(3)}_{ij}-4r^{-1}\dMbar^{(2)}_{ij}-r^{-2}\dMbar^{(1)}_{ij}-r^{-3}\dMbar_{ij}\bigg]\,.
	\end{align}
\end{subequations}
The tail terms in~\eqref{eq:hmunuMxMbarij} involve the modified Legendre function
\begin{equation}
	\label{eq:Qmbar}
	\overline{Q}_m(x,r) \equiv Q_m(x)-\frac{1}{2}P_m(x)\ln\left(\frac{r}{r_0}\right)\,,
\end{equation}
which is constructed using the usual Legendre polynomial $P_m(x)$ and the usual Legendre function of second kind (with branch cut \mbox{on $]-\infty,1]$)}, namely
\begin{equation}
	\label{eq:Qm}
	Q_m(x) \equiv \frac{1}{2}\int_{-1}^1 \dd y\,\frac{P_m(y)}{x-y} = \frac{1}{2} P_m(x) \ln \left(\frac{x+1}{x-1}\right)- \sum_{j=1}^m \frac{1}{m}P_{m-j}(x)P_{j-1}(x)\,.
\end{equation}
It was shown in~\cite{TLB22} that the combination~\eqref{eq:Qmbar} does not produce any far-zone logarithms in the radiative metric, although the tail term is still non-local in time and depends on the constant scale $r_0$. As shown in~\cite{TLB22} the metric~\eqref{eq:hmunuMxMbarij} differs from its counterpart in harmonic coordinates (given in Appendix~B in~\cite{BD92}) by a coordinate transformation and a redefinition of the quadrupole moment~\cite{TLB22}, see~\eqref{eq:momentRedefinition} below.

The radiative metric for the $\dS_i \times \dMbar_{ij}$ interaction is new with this paper and reads \cite{SuppMaterial}
\begin{subequations}\label{eq:hmunuSixMbarij}
	\begin{align}
	\label{subeq:h00SixMbarij}
		h_{\dS_i\times \dMbar_{ij}}^{00} &= n_{ad}\,\dS_{a\vert b}  \left(-\frac{4}{3} r^{-1} \dMbar_{bd}^{(4)} - 4 r^{-2} \dMbar_{bd}^{(3)}- 6 r^{-3}  \dMbar_{bd}^{(2)} -  6 r^{-4}\dMbar_{bd}^{(1)}  \right) \,,\\
	\label{subeq:h0iSixMbarij}
		h_{\dS_i\times \dMbar_{ij}}^{0i} &= n_{acd}\,\dS_{i\vert a}   \left(- \frac{5}{6} r^{-1} \dMbar_{cd}^{(4)}  - 5 r^{-2}   \dMbar_{cd}^{(3)}  - \frac{23}{2} r^{-3}   \dMbar_{cd}^{(2)} - \frac{3}{2} r^{-4}   \dMbar_{cd}^{(1)} - \frac{3}{2} r^{-5}   \dMbar_{cd}    \right) \nn\\
		& + n_{iad}\,\dS_{a\vert b}      \left(- 3 r^{-3} \dMbar_{bd}^{(2)} - 9 r^{-4} \dMbar_{bd}^{(1)} - 9 r^{-5} \dMbar_{bd} \right) \nn\\
		& + n_a\,\dS_{a\vert b}   \left( - \frac{4}{3}r^{-1}\dMbar_{ib}^{(4)} - \frac{4}{3}r ^{-2}\dMbar_{ib}^{(3)} + r^{-3}\dMbar_{ib}^{(2)} +  r^{-4}\dMbar_{ib}^{(1)} +  r^{-5}\dMbar_{ib} \right) \nn \\
		& +  n_a\,\dS_{i\vert b}    \left(  \frac{5}{3}r^{-2}\dMbar_{ab}^{(3)} +   5 r^{-3}\dMbar_{ab}^{(2)}  +   4 r^{-4}\dMbar_{ab}^{(1)} + 4 r^{-5}\dMbar_{ab}  \right) \,, \\
	\label{subeq:hijSixMbarij}
		h_{\dS_i\times \dMbar_{ij}}^{ij} &= n_{ijad}\,\dS_{a\vert b}  \left(\frac{1}{3}r^{-1}\dMbar_{bd}^{(4)}+\frac{10}{3}r^{-2}\dMbar_{bd}^{(3)}+9 r^{-3}\dMbar_{bd}^{(2)}+9r^{-4}\dMbar_{bd}^{(1)}\right)\nn\\
		&+n_{acd(i}\,\dS_{j)\vert a}   \left(\frac{1}{3}r^{-1}\dMbar_{cd}^{(4)}+\frac{10}{3}r^{-2}\dMbar_{cd}^{(3)}+9 r^{-3}\dMbar_{cd}^{(2)}+9 r^{-4}\dMbar_{cd}^{(1)}\right)\nn\\
		& + \delta_{ij} n_{ad}\,\dS_{a\vert b}   \left(r^{-1}\dMbar_{bd}^{(4)}+\frac{8}{3} r^{-2}\dMbar_{bd}^{(3)}-r^{-3}\dMbar_{bd}^{(2)}-r^{-4}\dMbar_{bd}^{(1)}\right)\nn\\
		& + n_{c(i}\,\dS_{j)\vert a}   \left(- \frac{4}{3} r^{-2}\dMbar_{ac}^{(3)}- 4 r^{-3}\dMbar_{ac}^{(2)}-4 r^{-4}\dMbar_{ac}^{(1)}\right)\nn\\
		& + n_{a(i}\,\dS_{a\vert b}  \left(- \frac{8}{3} r^{-1}\dMbar_{j)b}^{(4)} - \frac{26}{3} r^{-2}\dMbar_{j)b}^{(3)}-6  r^{-3}\dMbar_{j)b}^{(2)}-6 r^{-4}\dMbar_{j)b}^{(1)}\right)\nn\\
		& +  n_{ac}  \,\dS_{a\vert (i} \left(2r^{-1}\dMbar_{j)b}^{(4)}+ \frac{20}{3}r^{-2}\dMbar_{j)b}^{(3)}+ 2r^{-3}\dMbar_{j)b}^{(2)}+ 2r^{-4}\dMbar_{j)b}^{(1)}\right)\nn\\
		& + \,\dS_{a\vert (i} \left(- \frac{14}{3}r^{-2}\dMbar_{j)a}^{(3)} - 2 r^{-3}\dMbar_{j)a}^{(2)} - 2 r^{-4}\dMbar_{j)a}^{(1)}\right)\,.
	\end{align}
\end{subequations}
Note that contrary to the $\dM \times \dMbar_{ij}$ interaction in~\eqref{eq:hmunuMxMbarij}, the $\dS_i \times \dMbar_{ij}$ interaction is purely ``instantaneous'' or local, as it does not contain tail integrals. Note also that in the particular case of the $\dS_i \times \dMbar_{ij}$ interaction, the expressions for the radiative and harmonic metrics as functionals of their respective canonical moments are identical, and given by~\eqref{eq:hmunuSixMbarij}.

The radiative metric for the interaction $\dMbar_{ij}\times\dMbar_{ij}$ is too lengthy to be presented, so it is relegated to the Supplementary Material \cite{SuppMaterial}. It is constructed following Eq.~\eqref{eq:hnmunu} as
\begin{equation}\label{eq:hmunuMbarijxMbarij}
	h_{\dMbar_{ij}\times \dMbar_{ij}}^{\mu\nu} = u_{\dMbar_{ij}\times \dMbar_{ij}}^{\mu\nu} + v_{\dMbar_{ij}\times \dMbar_{ij}}^{\mu\nu} + \partial\xi_{\dMbar_{ij}\times \dMbar_{ij}}^{\mu\nu}\,.
\end{equation}
The gauge vector is defined by Eqs.~\eqref{eq:Lambdanmunu}--\eqref{eq:xinmu} and we have explicitly in this case~\cite{TLB22}
\begin{subequations}\label{eq:gaugeMbarijxMbarij}
\begin{align}\label{eq:sigmaMbarijxMbarij}
\sigma_{\dMbar_{ij} \times \dMbar_{ij}} 
= \hat{n}_{ijab} \,\dMbar_{ij}^{(3)}\dMbar_{ab}^{(3)} - \frac{24}{7} \hat{n}_{ij} \,\dMbar_{ia}^{(3)}\dMbar_{ja}^{(3)} + \frac{4}{5} \dMbar_{ab}^{(3)}\dMbar_{ab}^{(3)} \,,
\end{align}
and therefore
\begin{align}
		\label{subeq:xi0MbarijxMbarij}
		\xi_{\dMbar_{ij} \times \dMbar_{ij}}^0 &= \int_{-\infty}^{u} \dd v \!\int_1^{+\infty} \dd x \biggl\{- \frac{1}{2} \hat{n}_{ijab}\,Q_4(x) \dMbar_{ij}^{(3)}\dMbar_{ab}^{(3)} + \frac{12}{7} \hat{n}_{ij}\,Q_2(x) \dMbar_{ia}^{(3)}\dMbar_{ja}^{(3)} \nonumber \\
		&\qquad\qquad\qquad\qquad\quad - \frac{2}{5}\,Q_0(x) \dMbar_{ab}^{(3)}\dMbar_{ab}^{(3)}\biggr\}\,, \\
		\label{subeq:xiiMbarijxMbarij}
		\xi_{\dMbar_{ij} \times \dMbar_{ij}}^i &= \int_{-\infty}^{u} \dd v \!\int_1^{+\infty} \dd x \biggl\{ - \frac{1}{2} \hat{n}_{iabkl} \,Q_5(x) \dMbar_{ab}^{(3)}\dMbar_{kl}^{(3)} + \frac{16}{9} \hat{n}_{iab} \,Q_3(x) \dMbar_{ak}^{(3)}\dMbar_{kb}^{(3)} \nonumber \\
		&\qquad\qquad\qquad\qquad\quad - \frac{2}{9} \hat{n}_{abk} \,Q_3(x) \dMbar_{ai}^{(3)}\dMbar_{bk}^{(3)} - \frac{22}{35} \hat{n}_{i} \,Q_1(x) \dMbar_{ab}^{(3)}\dMbar_{ab}^{(3)} \nonumber \\
		&\qquad\qquad\qquad\qquad\quad + \frac{24}{35} \hat{n}_{a} \,Q_1(x) \dMbar_{ik}^{(3)}\dMbar_{ka}^{(3)} \biggr\}\,.
	\end{align}
\end{subequations}
Crucially, thanks to this gauge transformation the latter quadratic metric is free of any far-zone logarithms and so will be the cubic source built out of it. This allows us to take the dominant asymptotic behavior of the quadratic radiative metric as $r\to+\infty$, and extract after standard transverse-traceless (TT) projection of the spatial metric the associated radiative quadrupole moments (where we reintroduce $c$ and $G$):
\begin{subequations}
\label{eq:UijRadQuadratic}
\begin{align}
\label{subeq:UijRadMxMbarij}
\mathcal{U}_{ij}^{\dM\times \dMbar_{ij}} &= \frac{2 G \dM}{c^3} \int_0^{+\infty} \dd \tau\, \bigg[ \ln\left(\frac{\tau}{2r_0}\right) + \frac{107}{60}\bigg] \dMbar^{(4)}_{ij}(u-\tau)\,,\\
\label{subeq:UijRadSixMbarij}
\mathcal{U}_{ij}^{\dS_i \times \dMbar_{ij}} &= - \frac{G}{3 c^5}  \,\dS_{a\vert \langle i} \dMbar_{j\rangle a}^{(4)}\,,\\
\label{subeq:UijRadMbarijxMbarij}
\mathcal{U}_{ij}^{\dMbar_{ij}\times \dMbar_{ij}} &= -\frac{2G}{7c^5} \left[ \int_0^{+\infty} \dd \tau\,  \dMbar^{(3)}_{a \langle i}\dMbar^{(3)}_{j \rangle a}(u-\tau) + \dMbar^{(2)}_{a \langle i} \dMbar^{(3)}_{j \rangle a}+ \frac{5}{2} \dMbar^{(1)}_{a \langle i} \dMbar^{(4)}_{j \rangle a}- \frac{1}{2} \dMbar_{a \langle i} \dMbar^{(5)}_{j \rangle a}\right] \,.
\end{align}
\end{subequations}
The expression for the tail differs from (but is physically equivalent~\cite{TLB22} to)  its counterpart in the harmonic construction~\cite{BD92}; the expressions for $\dS_i\times \dMbar_{ij}$ and $\dMbar_{ij}\times \dMbar_{ij}$ are the same in both radiative and harmonic constructions, see the Appendix~B of~\cite{B98quad}. In particular the first term in~\eqref{subeq:UijRadMbarijxMbarij} is the usual memory effect (which contains both AC and DC contributions) at the dominant 2.5PN order.

\subsection{Structure of the cubic source}
\label{subsec:cubicSource}

The general equation we need to solve, Eq.~\eqref{eq:EinsteinEquationOrderN}, reads at cubic order
\begin{equation}
\label{eq:EinsteinEquationOrder3}
\Box h_3^{\mu\nu}- \partial H^{\mu\nu}_{3} = \Lambda^{\mu\nu}_3\bigl[h_1,h_2\bigr] = N^{\mu\nu}\bigl[h_1, h_2\bigr]+N^{\mu\nu}\bigl[h_2, h_1\bigr]+M^{\mu\nu}\bigl[h_1,h_1,h_1\bigr]\,,
\end{equation}
where we recall the definition $H_3^{\mu} \equiv \partial_\nu h_3^{\mu\nu}$. The quadratic and cubic functionals $N^{\mu\nu}_{\text{harm}}[h,h]$ and $M^{\mu\nu}_{\text{harm}}[h,h,h]$ in harmonic coordinates are given by (1.3)--(1.4) of~\cite{B98tail}, and to obtain their radiative counterparts $N^{\mu\nu}[h,h]$ and $M^{\mu\nu}[h,h,h]$ we apply~\eqref{eq:LambdaRad}, where the corrections terms are purely quadratic and thus only affect $N^{\mu\nu}[h,h]$. The source term for the cubic interaction $\dM\times\dMbar_{ij}\times\dMbar_{ij}$ reads, with obvious notation,
\begin{equation}
\Lambda^{\mu\nu}_{\dM \times \dMbar_{ij} \times \dMbar_{ij}} = N^{\mu\nu}_{\dMbar_{ij} \times (\dM \times \dMbar_{ij})} + N^{\mu\nu}_{\dM \times (\dMbar_{ij} \times \dMbar_{ij})} + M^{\mu\nu}_{\dM \times \dMbar_{ij} \times \dMbar_{ij}}\,.
\end{equation}
These three terms have the following structure:
\begin{itemize}
\item[1.] $\dMbar_{ij} \times (\dM \times \dMbar_{ij})$ corresponds to a linearized quadrupolar wave interacting with a quadratic tail, as illustrated by the first Feynman diagram in Fig.~\ref{subfig:Feynman1}. Apart from simple instantaneous terms, this source contains terms whose structure reads
\begin{align}\label{eq:complicatedStructureSource}
	\sim\frac{\hat{n}_L}{r^{5-p-q}}\,\dM \,\dMbar_{ab}^{(p)}(t-r) \int_1^{+\infty} \dd x \,\overline{Q}_m(x)\dMbar_{cd}^{(4+q)}(t-rx)\,, 
\end{align}
where $(p,q) \in [\![0,2 ]\!]$, and where the index structure and the possible presence of Kronecker deltas are disregarded. These terms are novel (with respect to ordinary tails), and their integration will be the focus of Sec.~\ref{subsec:ToMTechniques}. For integration purposes, the modified Legendre function $\overline{Q}_m(x)$ will be decomposed according to~\eqref{eq:Qmbar}: this will yield similar integrals over the standard Legendre function $Q_m(x)$, augmented by instantaneous terms involving $\ln\left(r/r_0\right)$.
\item[2.] $\dM \times (\dMbar_{ij} \times \dMbar_{ij})$ corresponds to a quadrupole-quadrupole memory-like wave scattering off the curvature generated by the total mass, and is illustrated by Fig.~\ref{subfig:Feynman2}. This source contains terms with the following structure:
\begin{subequations}
\begin{align}\sim\frac{\hat{n}_L}{r^{3-p-q}}\,\dM \int_1^{+\infty} \dd x\, \overline{Q}_m(x)\bigl(\dMbar_{ab}^{(3+p)}\,\dMbar_{cd}^{(3+q)}\bigr)(t-rx)\,,
\end{align}
where $0 \leqslant p+q \leqslant 2$, which have the same structure as ordinary tails, as well as memory-like terms
\begin{align}\sim\frac{\hat{n}_L}{r^4}\,\dM \int_0^{+\infty} \dd \tau\, \bigl(\dMbar_{ab}^{(3)}\,\dMbar_{cd}^{(3)}\bigr)(t-r-\tau)\,.
\end{align}
\end{subequations}
The modified Legendre function will again be decomposed according to~\eqref{eq:Qmbar}, and all these terms will be the treated in Sec.~\ref{subsec:TailsOfTailsAndTailsTechniques}.
\item[3.] $\dM \times \dMbar_{ij} \times \dMbar_{ij}$ properly speaking, illustrated by Fig.~\ref{subfig:Feynman3}. This piece only leads to instantaneous terms, which are easy to integrate.
\end{itemize}
A similar analysis can be done for the other two cubic interactions $\dM\times \dM \times \dMbar_{ij}$ and $\dM\times \dS_i \times \dMbar_{ij}$, which only contain tail-like and instantaneous terms. Note that, thanks to the radiative construction, we can see that the $r\to+\infty$ expansion of the cubic source does not contain any logarithms of $r$. We have explicitly verified that the novel source terms $\dM \times \dMbar_{ij} \times \dMbar_{ij}$ and $\dM \times \dS_{i} \times \dMbar_{ij}$ are divergenceless, \emph{i.e.}
\begin{equation}
	\partial_\nu\Lambda^{\mu\nu}_{\dM \times \dMbar_{ij} \times \dMbar_{ij}} = 0 \qquad \text{and} \qquad \partial_\nu\Lambda^{\mu\nu}_{\dM \times \dS_i \times \dMbar_{ij}} = 0\,,
\end{equation}
which is a strong test, since it would generally fail if a single coefficient were incorrect. However, because of its length, we cannot present the cubic source. Instead, we provide it in the form of an ancillary file in the Supplementary Material~\cite{SuppMaterial}.

\section{Solution of the wave equation at cubic order}
\label{sec:techniques}

\subsection{General multipolar solution}
\label{subsec:generalTechniques}

To solve the problem of tails-of-memory we need to develop new techniques for integrating the wave equation when the source term is a complicated non-local functional of the moments such as~\eqref{eq:complicatedStructureSource}. We first consider in a general way the wave equation whose source term admits a definite multipolarity $\ell$ in STF guise, thus takes the general form
\begin{equation}\label{eq:generalWaveEquation}
	\Box \Psi_L = \hat{n}_L\,S(r,t-r)\,,
\end{equation}
where $S(r,u)$ is an arbitrary function of $r=\vert\mathbf{x}\vert$ and $u=t-r$ that verifies straightforward smoothness properties, and tends sufficiently rapidly to zero when $r\to 0$, namely
\begin{equation}\label{eq:falloffNZ}
	S(r,u) = \mathcal{O}(r^{\ell+5})\quad\text{(when $r\to 0$ with $u$ or $t$ held fixed)}\,.
\end{equation}
See Theorem 6.1 of~\cite{BD86} for more details on the required conditions we impose. Under these conditions, we know how to solve the wave equation~\eqref{eq:generalWaveEquation}. We first define
\begin{equation}\label{eq:Ralpha}
	R_\alpha(\rho, s) \equiv \rho^\ell \int_\alpha^\rho \dd\lambda\, \frac{(\rho-\lambda)^\ell}{\ell!}\left(\frac{2}{\lambda}\right)^{\ell-1} S(\lambda,s) \,,
\end{equation}
where $\alpha$ is an arbitrary constant. Then the solution of~\eqref{eq:generalWaveEquation} can be written as
\begin{equation}\label{eq:PsiLwithRalpha} \Psi_L = \int_{-\infty}^{t-r}\dd s\, \hat{\partial}_L \left[ \frac{R_\alpha\left(\frac{t-r-s}{2}, s\right)-R_\alpha\left(\frac{t+r-s}{2}, s\right)}{r}\right]\,,
\end{equation}
see Eq.~(6.4) in~\cite{BD86}. This solution is the unique retarded solution of~\eqref{eq:generalWaveEquation}, thus satisfying the no-incoming radiation condition when $r\to+\infty$ with $t+r=$ const. Furthermore it is independent from the constant $\alpha$. To see this we consider separately the two terms in~\eqref{eq:PsiLwithRalpha}, which we decompose as \mbox{$\Psi_L=\Psi^{[1]}_L+\Psi^{[2]}_L$} with
\begin{subequations}\label{eq:PsiLABwithRalpha}
	\begin{align}
	\label{eq:PsiLAwithRalpha}
		\Psi^{[1]}_L &\equiv  \int_{-\infty}^{t-r}\dd s\, \hat{\partial}_L \left[ \frac{1}{r}R_\alpha\left(\frac{t-r-s}{2}, s\right)\right] \,,\\
	\label{eq:PsiLBwithRalpha}
		\Psi^{[2]}_L &\equiv - \int_{-\infty}^{t-r}\dd s\, \hat{\partial}_L \left[ \frac{1}{r}R_\alpha\left(\frac{t+r-s}{2}, s\right)\right] \,.
	\end{align}
\end{subequations}
We see that $\Psi^{[1]}_L$ is a retarded homogeneous solution of the wave equation, $\Box\Psi^{[1]}_L=0$, but that $\Psi^{[2]}_L$ has a unwieldy dependency on the advanced time $v=t+r$, and represents a particular retarded solution of the wave equation, \emph{i.e.}, $\Box\Psi^{[2]}_L=\hat{n}_L\,S(r,t-r)$. However note that the latter two wave equations satisfied by $\Psi^{[1]}_L$ and $\Psi^{[2]}_L$ separately are valid only in the sense of ordinary functions; only the wave equation~\eqref{eq:generalWaveEquation} for the full solution $\Psi_L$ is satisfied in the sense of distributions.\footnote{If $f$ is a smooth function, then $\Box \left[r^{-1} f(t-r)\right] = -4\pi f(t) \delta^{(3)}(\mathbf{x})$ in the sense of distributions, but $\Box \left[r^{-1} f(t-r)\right] = 0$ in the sense of ordinary functions.}
\newpage
Plugging Eq.~\eqref{eq:Ralpha} into~\eqref{eq:PsiLABwithRalpha} we obtain
\begin{subequations}\label{eq:PsiLABwithS}
	\begin{align}
	\label{eq:PsiLAwithS}
		\Psi^{[1]}_L &= \frac{1}{\ell!} \int_{-\infty}^{t-r}\dd s\, \hat{\partial}_L\!\left[ \frac{1}{r}\int_\alpha^{\frac{t-r-s}{2}} \!\!\dd \lambda \left(\frac{t-r-s}{2}\right)^\ell \!\left(\frac{t-r-s}{2}-\lambda\right)^\ell \left(\frac{2}{\lambda}\right)^{\ell-1} \!\!S(\lambda, s)\right]\,,\\
	\label{eq:PsiLBwithS}
		\Psi^{[2]}_L &= - \frac{1}{\ell!} \int_{-\infty}^{t-r}\dd s\, \hat{\partial}_L\!\left[ \frac{1}{r}\int_\alpha^{\frac{t+r-s}{2}} \!\!\dd \lambda \left(\frac{t+r-s}{2}\right)^\ell \!\left(\frac{t+r-s}{2}-\lambda\right)^\ell \left(\frac{2}{\lambda}\right)^{\ell-1}\!\!S(\lambda, s)\right]\,.
	\end{align}
\end{subequations}
Next we remark that in both terms~\eqref{eq:PsiLABwithS}, one can commute the partial differential \mbox{operator $\hat{\partial}_L$} with the integral over $\lambda$, since the terms coming from the differentiation of the bound of the integral, \emph{i.e.} $(t\pm r-s)/2$, have to be evaluated at $s=t\pm r$ and will clearly vanish. Hence we can rewrite
\begin{subequations}\label{eq:PsiLABwithSv2}
	\begin{align}
	\label{eq:PsiLAwithSv2}
		\Psi^{[1]}_L &= \frac{2^{\ell-1}}{\ell!} \int_{-\infty}^{t-r}\dd s\!\int_\alpha^{\frac{t-r-s}{2}} \!\!\dd \lambda \, \lambda^{-\ell+1} S(\lambda, s) \,\hat{\partial}_L\!\left[\frac{1}{r}\left(\frac{t-r-s}{2}\right)^\ell \,\left(\frac{t-r-s}{2}-\lambda\right)^\ell \right]\,,\\
	\label{eq:PsiLBwithSv2}
		\Psi^{[2]}_L &= - \frac{2^{\ell-1}}{\ell!} \int_{-\infty}^{t-r}\dd s\!\int_\alpha^{\frac{t+r-s}{2}} \!\!\dd \lambda \, \lambda^{-\ell+1} S(\lambda, s) \,\hat{\partial}_L\!\left[\frac{1}{r}\left(\frac{t+r-s}{2}\right)^\ell \!\left(\frac{t+r-s}{2}-\lambda\right)^\ell \right]\,.
	\end{align}
\end{subequations}
Then, using Eq.~(A36) of~\cite{BD86} we know that
\begin{equation}\label{eq:formulaRet2AdvTime}
	\hat{\partial}_L\!\left[\frac{1}{r}\left(\frac{t-r-s}{2}\right)^\ell\left(\frac{t-r-s}{2}-\lambda\right)^\ell\right]= \hat{\partial}_L\!\left[\frac{1}{r}\left(\frac{t+r-s}{2}\right)^\ell\left(\frac{t+r-s}{2}-\lambda\right)^\ell\right]\,,
\end{equation}
so that actually the two contributions in~\eqref{eq:PsiLABwithSv2} can be merged together in the full solution, and we arrive at the elegant alternative form 
\begin{equation}\label{eq:PsiLwithS}
	\Psi_L = - \frac{2^{\ell-1}}{\ell!} \int_{-\infty}^{t-r}\dd s\!\int_{\frac{t-r-s}{2}}^{\frac{t+r-s}{2}} \!\!\dd \lambda \, \lambda^{-\ell+1} S(\lambda, s) \,\hat{\partial}_L\!\left[\frac{1}{r}\left(\frac{t-r-s}{2}\right)^\ell \,\left(\frac{t-r-s}{2}-\lambda\right)^\ell \right]\,.
\end{equation}
This form is clearly independent of any choice for the constant $\alpha$. It was also derived in a different way, using a direct multipolar expansion of the Green function of the d'Alembertian operator, in Appendix~D of~\cite{BD86}.

By using Eq.~(A35a) of~\cite{BD86} we can expand explicitly the last factor containing the multipolar derivative operator. Introducing the following coefficients, which will become ubiquitous in our practical computations,
\begin{equation}\label{eq:Cij}
	C_{ij}^\ell\equiv\frac{(\ell+i)!(\ell+j)!}{2^{i+j}i!j!(\ell-i)!(\ell-j)!(i+j)!}\\\qquad\text{(for $0\leqslant i,j\leqslant\ell$)}\,,
\end{equation}
we conveniently write the general solution as
\begin{equation}\label{eq:PsiLwithSv2}
	\Psi_L =  - \frac{\hat{n}_L}{2}\sum_{i=0}^\ell \sum_{j=0}^\ell \,\frac{(-)^j  C_{ij}^\ell }{r^{i+1}} \int_{0}^{+\infty}\dd \rho\, \rho^{i+j} \int_{ \frac{\rho}{2}}^{ \frac{\rho}{2}+r}\dd\lambda\,\lambda^{-j+1} \,S(\lambda, u-\rho) \,.
\end{equation}
Note that although the coefficient $C_{ij}^\ell$ has been defined to be symmetric in $i$ and $j$, these two indices play a different role. In particular the index $i$ rules the behavior of the solution at infinity, when $r\to+\infty$ with $t-r=$ const, depending on the fall-off properties at infinity of the integrals in~\eqref{eq:PsiLwithSv2}. A useful fact is that for $i=0$, the coefficient is straightforwardly linked to the $j$-th derivative of the usual Legendre polynomial evaluated at 1, namely
\begin{equation}\label{eq:C0j}
	C_{0j}^\ell = \frac{1}{j!}P_\ell^{(j)}(1)\,.
\end{equation}

In addition to the general formula~\eqref{eq:PsiLwithSv2}, we also have an independent result which directly provides the leading behavior of the solution at infinity, when $r\to+\infty$ with \mbox{$u=$ const}, depending on the fall-off properties of the source $S(r,u)$. Namely, suppose that $S(r,u)$ has the following asymptotic behavior at infinity:
\begin{equation}\label{eq:falloffFZ}
	S(r,u) = \mathcal{O}\left(\frac{1}{r^3}\right)\quad\text{(when $r\to +\infty$ with $u$ fixed)}\,.
\end{equation}
More precisely the rigourous fall-off conditions of the source term are given in Lemma 7.2 of~\cite{BD86}. Then the corresponding solution will behave dominantly like $1/r$ at infinity, with leading term explicitly given by
\begin{equation}\label{eq:PsiLwithRalphaAsympt}
	\Psi_L = \frac{(-)^\ell}{2 ^\ell} \frac{\hat{n}_L}{r} \int_{-\infty}^{t-r}\dd s\, R_{\infty}^{(\ell)}\left(\frac{t-r-s}{2}, s\right) + \mathcal{O}\left(\frac{1}{r^2}\right)\,.
\end{equation}
Here the function $R_{\infty}(\rho,s)$ is defined by~\eqref{eq:Ralpha} with the explicit choice $\alpha=+\infty$, and the superscript $(\ell)$ means $\ell$ partial derivatives with respect to $\rho$. We shall use the latter result as a consistency check of the derivation of the leading behavior of the solution at infinity.  

\subsection{Application to tails-of-memory}
\label{subsec:ToMTechniques}

We apply the general formalism in the previous section to the cubic iteration and specifically to the tails-of-memory. In this case the main problem we face is to find the solution of the wave equation~\eqref{eq:generalWaveEquation} when the source term takes the form~\eqref{eq:complicatedStructureSource}. For ease of notation we denote the two quadrupole moments by arbitrary time-varying functions $F(u)$ and $G(u)$, and consider the source term (multiplying the STF harmonics $\hat{n}_L$):
\begin{equation}\label{eq:sourceBToM}
	\underset{k,m}{\ \ S^B}(r,t-r) \equiv \left(\frac{r}{r_0}\right)^B r^{-k}\,G(t-r)\int_1^{+\infty} \dd x\, Q_m(x) F(t-rx)\,,
\end{equation}
where $k\geqslant 1$ and $m\geqslant 0$ are integers, and the Legendre function of the second kind $Q_m(x)$ is given by~\eqref{eq:Qm}, as was obtained from $\overline{Q}_m(x)$ using the relation \eqref{eq:Qmbar}. The source~\eqref{eq:sourceBToM} is more complex than the one required for the computation of tails and tails-of-tails. In the latter case, the function $G(u)$ is actually constant (\emph{i.e.} the ADM mass $\dM$), and we shall recover and extend the results found in~\cite{B98tail, MBF16}. In the case where $G(u)$ and $F(u)$ are not constant the results we shall derive are new. 

In~\eqref{eq:sourceBToM} we have multiplied the source by the regularization factor $(r/r_0)^B$, where $B\in\mathbb{C}$ and $r_0$ is an arbitrary constant, see Eq.~\eqref{eq:unMPM}. Very importantly, this permits to ensure after analytic continuation in $B\in\mathbb{C}$ that the fall-off condition~\eqref{eq:falloffNZ} is always satisfied. Applying the finite part, the solution we are looking for is
\begin{equation}\label{eq:defPsikmL}
	 \underset{k,m}{\ \Psi_L} \equiv \pf \underset{k,m}{\ \Psi^B_L}\,,\qquad \text{where} \qquad \underset{k,m}{\ \Psi^B_L} \equiv \Box_\text{ret}^{-1} \!\underset{k,m}{\ \ S^B_L} = \Box_\text{ret}^{-1} \bigl[\hat{n}_L\!\!\underset{k,m}{\ \ S^B}\bigr]\,,
\end{equation}
and will be obtained by applying the formulae of the previous section. During the computation, we shall also encounter explicit logarithms and poles proportional to $1/B$. This will all boil down to computing 
\begin{equation}\label{eq:defChikmL}
 \underset{k,m}{\ \chi_L} \equiv \pf \biggl[ \frac{1}{B} \underset{k,m}{\ \Psi^B_L} \biggr] = \pf \left[ \frac{\dd}{\dd B}\!\!\underset{k,m}{\ \ \Psi^B_L} \right] = \FPprop \left[  \ln\left(\frac{r}{r_0}\right)\!\!\underset{k,m}{\ \ S^B_L} \right]\,,
\end{equation}
where we have used the useful property of the regularization factor $(r/r_0)^B$, that an extra logarithm is generated by differentiating with respect to $B$:
\begin{equation}\label{eq:identitySkmBDiffLog}
	\frac{\dd}{\dd B}\!\underset{k,m}{\ \ S^B_L} = \ln\left(\frac{r}{r_0}\right)\!\!\underset{k,m}{\ \ S^B_L}\,.
\end{equation}
Applying the formulae of the previous section the solution for any $B \in \mathbb{C}$ reads
\begin{subequations}
\label{eq:expressionPsiBkml} 
\begin{align}\label{subeq:expressionPsiBkmlwithPhiBkmij} 
	\underset{k,m}{\ \Psi^B_L} &=  - \frac{\hat{n}_L}{2}\sum_{i=0}^\ell \sum_{j=0}^\ell \frac{(-)^j C_{ij}^\ell}{r^{i+1}}\underset{k,m}{\phi}\!_{ij}^B \,,
\end{align}
where
\begin{align}\label{subeq:expressionPhiBkmij} 
	\underset{k,m}{\phi}\!_{ij}^B &\equiv  \int_{0}^{+\infty}\!\!\dd \rho\, \rho^{i+j}G(u-\rho) \int_{ \frac{\rho}{2}}^{ \frac{\rho}{2}+r}\!\!\dd\lambda\, \left(\frac{\lambda}{r_0}\right)^B \lambda^{-k-j+1}  \int_1^{+\infty} \!\!\dd x \, Q_m(x) F\bigl[u-\rho-\lambda(x-1)\bigr] \,.
\end{align}
\end{subequations}
Note that this solution is ``exact'', valid at any radial distance $r$ except $r=0$. But in the following, we shall mostly be interested in the asymptotic limit when $r\to+\infty$ with \mbox{$u=$ const}. Permuting the integrals, introducing the change of variable $\lambda\longrightarrow \tau = \lambda (x-1)$, and factorizing out the expected leading behavior $1/r$ of the solution (in anticipation of the limit $r\to+\infty$), we find that
\begin{subequations}\label{eq:expressionPsiBkmlwithKernelKB}
	\begin{align}  
		\label{subeq:expressionPsiBkmlwithKernelKB}
		\underset{k,m}{\ \Psi^B_L} &=  - \frac{\hat{n}_L}{2r} \int_{0}^{+\infty} \dd \rho\, G(u-\rho) 
		\int_{0}^{+\infty} \dd\tau \, F(u-\rho-\tau) \underset{k,m}{\ K^B_\ell}(\rho,\tau,r) \,,
	\end{align}
where we have introduced for convenience the kernel function (with $x=y+1$)
\begin{align}  
	\label{subeq:expressionKernelKB}
	\underset{k,m}{\ K^B_\ell}(\rho,\tau,r) &=  \left(\frac{\tau}{r_0}\right)^{B}\sum_{i=0}^\ell \frac{1}{r^i}\sum_{j=0}^\ell (-)^j C_{ij}^\ell \,\rho^{i+j}\,\tau^{-k-j+1} \int_{\frac{2\tau}{\rho+2r}}^{\frac{2\tau}{\rho}} \dd y \, y^{k+j-2-B} Q_m(y+1) \,,\nn\\
	&= \tau^{1-k}\int_{\frac{2\tau}{\rho+2r}}^{\frac{2\tau}{\rho}}\dd y\, \left(\frac{\tau}{y r_0}\right)^{B} y^{k-2}Q_m(y+1)\,\Pi_\ell\left(1- \frac{\rho y}{\tau}, 1+\frac{\rho}{r}\right)\,.
\end{align}
\end{subequations}
\newpage
In the second line we have introduced the following symmetric bivariate polynomial 
\begin{equation}
	\label{eq:Pi}
	\Pi_\ell(x,y) = \sum_{i=0}^\ell \sum_{j=0}^\ell C^\ell_{ij} (x-1)^i (y-1)^j\,,
\end{equation}
which is related to the Legendre polynomial, recalling~\eqref{eq:C0j}, by  
\begin{equation}
\label{eq:relationPiandP}
P_\ell(x) = \sum_{j=0}^\ell  C^\ell_{0j}\left(x-1\right)^j = \Pi_\ell(x,1) =  \Pi_\ell(1,x) \,.
\end{equation}
Similarly, the quantity $_{k,m}\chi_L^B$ defined in~\eqref{eq:defChikmL} reads
\begin{subequations}\label{eq:expressionChiBkmlwithKernelLB}
	\begin{align}  
		\label{subeq:expressionChiBkmlwithKernelLB}
		\underset{k,m}{\ \chi^B_L} &=  - \frac{\hat{n}_L}{2r} \int_{0}^{+\infty} \dd \rho\, G(u-\rho) 
		\int_{0}^{+\infty} \dd\tau \, F(u-\rho-\tau) \underset{k,m}{\ L^B_\ell}(\rho,\tau,r) \,,
\end{align}	
where the kernel is defined as
\begin{align}
	\label{subeq:expressionKernelLB}
	\underset{k,m}{\ L^B_\ell} &= \frac{\dd}{\dd B}\biggl[ \underset{k,m}{\ K^B_\ell}(\rho,\tau,r) \biggr] \nn\\
	 &= \sum_{i=0}^\ell \frac{1}{r^i}\sum_{j=0}^\ell (-)^j C_{i,j}^\ell \,\rho^{i+j}\,\tau^{-k-j+1} \int_{\frac{2\tau}{\rho+2r}}^{\frac{2\tau}{\rho}} \dd y \left(\frac{\tau}{y r_0}\right)^{B} \,\ln\left(\frac{\tau}{r_0 y}\right) y^{k+j-2} Q_m(x) \,, \nn\\
	&= \tau^{1-k}\int_{\frac{2\tau}{\rho+2r}}^{\frac{2\tau}{\rho}}\dd y\, \left(\frac{\tau}{y r_0}\right)^{B} \ln\left(\frac{\tau}{y r_0}\right) y^{k-2}Q_m(y+1)\,\Pi_\ell\left(1- \frac{\rho y}{\tau}, 1+\frac{\rho}{r}\right)\,.
\end{align}
\end{subequations}

When $r\to +\infty$ the above kernels are dominated by the contribution $i=0$, and the property~\eqref{eq:relationPiandP} allows us to relate the asymptotic limit to the Legendre polynomial as
\begin{subequations}\label{eq:KernelsBAsymptotic}
	\begin{align} 
		\underset{k,m}{\ K^B_\ell} &=  \tau^{1-k}\int_{\frac{2\tau}{\rho+2r}}^{\frac{2\tau}{\rho}}\dd y\, \left(\frac{\tau}{y r_0}\right)^{B} y^{k-2}Q_m(y+1)\,P_\ell\left(1- \frac{\rho y}{\tau}\right) + o\left(1\right) \,,\\
		\underset{k,m}{\ L^B_\ell}&=  \tau^{1-k}\int_{\frac{2\tau}{\rho+2r}}^{\frac{2\tau}{\rho}}\dd y\, \left(\frac{\tau}{y r_0}\right)^{B} \ln\left(\frac{\tau}{y r_0}\right)  y^{k-2}Q_m(y+1)\,P_\ell\left(1- \frac{\rho y}{\tau}\right) + o\left(1\right) \,.
	\end{align}
\end{subequations}
We employ the Landau symbol $o$ for remainders, hence $o(1)$ means terms that behave as $\sim \ln^p r/r$ with uncontrolled powers of $\ln r$ as $r\rightarrow +\infty$.

\subsubsection{Cases where $k=1$ and $k=2$}
\label{subsubsec:ToMTechniquesCasek12}

In this subsection, we assume that $k\in\{1,2\}$. As proven in Appendix~\ref{app:convergence}, $_{k,m}\, \Psi^B_L$ and $_{k,m}\,\chi^B_L$ have a well-defined limit when $B\rightarrow 0$, so we can drop the finite part prescription and simply set $B=0$ in~\eqref{eq:expressionPsiBkmlwithKernelKB} and~\eqref{eq:expressionChiBkmlwithKernelLB}. Therefore we obtain the expression for the kernels, only valid for $k\in\{1,2\}$, 
\newpage
\begin{subequations}
\label{eq:defKernelsKL}
\begin{align}
\label{subeq:defKernelK}
	\underset{k,m}{\ K_{\ell}}(\rho,\tau,r) &\equiv \tau^{1-k}\int_{\frac{2\tau}{\rho+2r}}^{\frac{2\tau}{\rho}}\dd y\, y^{k-2}Q_m(y+1)\,\Pi_\ell\left(1- \frac{\rho y}{\tau}, 1+\frac{\rho}{r}\right)\, , \\
\label{subeq:defKernelL}
	\underset{k,m}{\ L_{\ell}}(\rho,\tau,r) &\equiv \tau^{1-k}\int_{\frac{2\tau}{\rho+2r}}^{\frac{2\tau}{\rho}}\dd y\, \ln\left(\frac{\tau}{y r_0}\right)y^{k-2}Q_m(y+1)\,\Pi_\ell\left(1- \frac{\rho y}{\tau}, 1+\frac{\rho}{r}\right)\,.
\end{align}
\end{subequations}
When restricting our interest to the asympotic limit $r \rightarrow +\infty$, we must treat separately the cases $k=1$ and $k=2$.

For $k=2$, the integrands in the kernels~\eqref{eq:defKernelsKL} are clearly integrable at the $y\rightarrow 0$ bound, therefore the kernels converge in the $r \rightarrow +\infty$ limit, such that we are allowed to define the ``asymptotic'' kernels:
\begin{subequations}\label{eq:defAsympKernelsKbarLbar}
	\begin{align}  
	\label{subeq:defAsympKernelKbar}
		\underset{2,m}{\ \overline{K}_\ell}(\rho,\tau) &\equiv \lim_{r\to+\infty} \underset{2,m} {\ K_\ell}(\rho,\tau,r) = \tau^{-1} \int_{0}^{\frac{2\tau}{\rho}} \dd y \,Q_m(y+1) P_\ell\Bigl(1-\frac{\rho y}{\tau}\Bigr)\,,\\
	\label{subeq:defAsympKernelLbar}
		\underset{2,m}{\ \overline{L}_\ell}(\rho,\tau) & \equiv \lim_{r\to+\infty} \underset{2,m} {\ L_\ell}(\rho,\tau,r) = \tau^{-1} \int_{0}^{\frac{2\tau}{\rho}} \dd y \,\ln\left(\frac{\tau}{y r_0 }\right) Q_m(y+1) P_\ell\Bigl(1-\frac{\rho y}{\tau}\Bigr)\,.
	\end{align}
\end{subequations}
We have verified the above asymptotic limit in the case where $k=2$ using the general statement in Eq.~\eqref{eq:PsiLwithRalphaAsympt}, where the relevant function is defined by Eq.~\eqref{eq:Ralpha} with $\alpha=+\infty$. 

In the case $k=1$, the situation is more complicated since the integrands of the kernels are no longer integrable when $y\to 0$, so it is not possible to simply take the limit $r\to+\infty$ like in~\eqref{eq:defAsympKernelsKbarLbar}. Instead the kernels exhibit a logarithmic behavior as $r\to+\infty$. Since the radiative construction will not exhibit any logarithmic behavior, the logarithms should cancel out in the final metric, and it is crucial to verify this by controlling the logarithmic limit of the kernel functions. We detail the case of the ``$K$''-kernel
\begin{equation}
\label{eq:defKernelK1}
	\underset{1,m}{\ K_{\ell}}(\rho,\tau,r) = \int_{\frac{2\tau}{\rho+2r}}^{\frac{2\tau}{\rho}}\frac{\dd y}{y} \,Q_m(y+1)\,\Pi_\ell\left(1- \frac{\rho y}{\tau}, 1+\frac{\rho}{r}\right)\,.
\end{equation}
To extract the logarithmic behavior, we integrate by parts. First, we introduce the regular part of the Legendre function of the second type when $x\to 1^+$, which is defined as
\begin{equation}
\label{eq:Rm}
R_m(x) \equiv Q_m(x) + \frac{1}{2} P_m(x) \ln\left(\frac{x-1}{2}\right)\,.
\end{equation}
A useful observation is that $R_m(1) = -H_m$, where $H_m=\sum_{k=1}^{m}k^{-1}$ is the harmonic number. Substituting $Q_m$ by its expression in terms of $R_m$ and  integrating by parts, we find
\begin{align}
\label{eq:expressionKernelK1IBP}
	\underset{1,m}{\ K_{\ell}}(\rho,\tau,r) &=\Bigg[\ln\left(\frac{y}{2}\right)\biggl(R_m(y+1)-\frac{1}{4}\ln\left(\frac{y}{2}\right)P_m(y+1)\biggr)\Pi_\ell\left(1-\frac{\rho y}{\tau},1+\frac{\rho}{r}\right)\Bigg]_{y=\frac{2\tau}{\rho+2r}}^{y=\frac{2\tau}{\rho}}\nn\\
	&-\int_{\frac{2\tau}{\rho+2r}}^{\frac{2\tau}{\rho}}\dd y\, \ln\left(\frac{y}{2}\right)\frac{\dd }{\dd y}\bigg[ R_m(y+1)\,\Pi_\ell\left(1- \frac{\rho y}{\tau}, 1+\frac{\rho}{r}\right)\bigg]\nn\\
	&+\frac{1}{4}\int_{\frac{2\tau}{\rho+2r}}^{\frac{2\tau}{\rho}}\dd y\, \ln^2\left(\frac{y}{2}\right)\frac{\dd }{\dd y}\bigg[P_m(y+1)\,\Pi_\ell\left(1- \frac{\rho y}{\tau}, 1+\frac{\rho}{r}\right)\bigg] \,,
\end{align}
where the all-integrated terms are shown in the first line. Since $P_m(x)$, $R_m(x)$ and $\Pi_\ell(x,y)$ as well as all their derivatives are perfectly integrable as $x\rightarrow 1^+$, we can safely take the $r\to+\infty$ expansion, and we obtain the explicit ``polylogarithmic'' structure
\begin{subequations}
\label{eq:expressionKernelK1Asymp}
\begin{align}\label{subeq:expressionKernelK1Asymp}
	\underset{1,m}{\ K_{\ell}}(\rho,\tau,r) &= \frac{1}{4}\ln^2\left(\frac{r}{r_0}\right)-\frac{1}{2}\ln\left(\frac{r}{r_0}\right)\bigg[\ln\left(\frac{\tau}{2r_0}\right)+2 H_m\bigg]+ \underset{1,m}{\ \overline{K}_{\ell}}(\rho,\tau) + o\left(1\right)\,,
\end{align}
where ${}_{1,m}\overline{K}_{\ell}(\rho,\tau)$ does not exhibit any $r$-dependence. We obtain its explicit expression as
\begin{align}\label{subeq:expressionK1bar}
	\underset{1,m}{\ \overline{K}_{\ell}}(\rho,\tau) &= \frac{1}{4}\ln^2\left(\frac{\tau}{2r_0}\right)+ H_m \ln\left(\frac{\tau}{2r_0}\right)\nn\\
	& - \frac{(-)^\ell}{4} \left[\ln^2 \left(\frac{\tau}{2 r_0}\right) -2 \ln \left(\frac{\tau}{2 r_0}\right) \ln \left(\frac{\rho}{2 r_0}\right) + \ln^2 \left(\frac{\rho}{2 r_0}\right)\right]P_m\left(1+ \frac{2\tau}{\rho}\right) \nn\\
	& - (-)^\ell\left[ \ln \left(\frac{\rho}{2 r_0}\right)-  \ln \left(\frac{\tau}{2 r_0}\right)\right]R_m\left(1+ \frac{2\tau}{\rho}\right) \nn\\
	&+\frac{1}{4}\int_{0}^{\frac{2\tau}{\rho}}\dd y\, \ln^2\left(\frac{y}{2}\right)\frac{\dd }{\dd y}\bigg[P_m(y+1)\,P_\ell\left(1- \frac{\rho y}{\tau}\right)\bigg] \nn\\
	&-\int_{0}^{\frac{2\tau}{\rho}}\dd y\, \ln\left(\frac{y}{2}\right)\frac{\dd }{\dd y}\bigg[ R_m(y+1)\,P_\ell\left(1- \frac{\rho y}{\tau}\right)\bigg] \,.
\end{align}
\end{subequations}
The same reasoning applied to the ``$L$''-kernel also gives a polylogarithmic structure but in this case with powers of the logarithm up to three:
\begin{subequations}
\label{eq:expressionKernelL1Asymp}
\begin{align}\label{subeq:expressionKernelL1Asymp}
	\underset{1,m}{\ L_{\ell}}(\rho,\tau,r) &= \frac{1}{6} \ln^3\left(\frac{r}{r_0}\right)-\frac{1}{4}\ln^2\left(\frac{r}{r_0}\right)\left[\ln\left(\frac{\tau}{2r_0}\right)+2 H_m\right]+ \underset{1,m}{\ \overline{L}_{\ell}}(\rho,\tau)  + o\left(1\right) \,,
\end{align}
and where ${}_{1,m}\overline{L}_{\ell}(\rho,\tau)$ is explicitly given by
\begin{align}\label{subeq:expressionL1bar}
	\underset{1,m}{\ \overline{L}_{\ell}}(\rho,\tau) &= \frac{1}{12}\ln^3\left(\frac{\tau}{2r_0}\right)+ \frac{1}{2}H_m \ln^2\left(\frac{\tau}{2r_0}\right)\\
	& - \frac{(-)^\ell}{12} \left[\ln^3 \left(\frac{\tau}{2 r_0}\right) -3 \ln \left(\frac{\tau}{2 r_0}\right) \ln^2 \left(\frac{\rho}{2 r_0}\right) + 2\ln^3 \left(\frac{\rho}{2 r_0}\right)\right]P_m\left(1+ \frac{2\tau}{\rho}\right) \nn\\
	& - \frac{(-)^\ell}{2}\left[ \ln^2 \left(\frac{\rho}{2 r_0}\right)-  \ln^2 \left(\frac{\tau}{2 r_0}\right)\right]R_m\left(1+ \frac{2\tau}{\rho}\right)\nn \\
	&+\int_{0}^{\frac{2\tau}{\rho}}\dd y\,\ln^2\left(\frac{y}{2}\right) \left[\frac{1}{4}\ln\left(\frac{\tau}{2r_0}\right)-\frac{1}{6}\ln\left(\frac{y}{2}\right)\right]\frac{\dd }{\dd y}\bigg[P_m(y+1)\,P_\ell\left(1- \frac{\rho y}{\tau}\right)\bigg] \nn\\
	&+\int_{0}^{\frac{2\tau}{\rho}}\dd y\,\ln\left(\frac{y}{2}\right) \left[-\ln\left(\frac{\tau}{2r_0}\right)+\frac{1}{2}\ln\left(\frac{y}{2}\right)\right]\frac{\dd }{\dd y}\bigg[R_m(y+1)\,P_\ell\left(1- \frac{\rho y}{\tau}\right)\bigg] \,.\nn
\end{align}
\end{subequations}

Again, a beautiful check of the radiative construction of the metric in Sec.~\ref{sec:radMPM}, is that all these explicitly determined far zone logarithms will be compensated by those induced by the applied gauge transformations, notably the one described in~\eqref{eq:gaugeMbarijxMbarij}.

In practical computations, it is easier to compute the full kernels ${}_{1,m}K_\ell(\rho, \tau, r)$ and ${}_{1,m}L_\ell(\rho, \tau, r)$, and then to remove the logarithmic dependencies to obtain the reduced asymptotic kernels ${}_{1,m}\overline{K}_\ell(\rho, \tau)$ and ${}_{1,m}\overline{L}_\ell(\rho, \tau)$. However, computing the two later quantities directly from their explicit expressions~\eqref{subeq:expressionK1bar} and~\eqref{subeq:expressionL1bar} yields, of course, the same result.

\subsubsection{Cases where $k\geqslant 3$}
\label{subsubsec:ToMTechniquesCasek3}

In the previous subsection we could integrate the source term~\eqref{eq:sourceBToM} when $k\in\{1,2\}$. \emph{A priori}, the cases $k \geqslant 3$ are more difficult and (when $G$ is not constant) analytic closed-form expressions for the retarded integral seem non trivial. Hence we proceed differently and prove that we can retrieve the cases $k \geqslant 3$ from the known cases $k\in\{1,2\}$. This will show that analytic closed-form expressions also exist in the cases $k\geqslant 3$.

Given a source term with $k\geqslant 3$, endowed with the associated regularization factor $r^B$ (posing here $r_0=1$), we reduce it to a fully integrated part and new source terms with decreased values of $k$ by two steps at most: $k-1$ and $k-2$, by means of the identity:
\begin{align}\label{eq:kReducingIdentityToM}
	&\hat{n}_L r^{B-k} G(t-r) \int_1^{+\infty} \dd x Q_m(x) F(t-rx)\nn \\
	& \qquad = \Box\left[ \frac{\hat{n}_L r^{B-k+2}}{(k+\ell-2-B)(k-\ell-3-B)} G(t-r) \int_1^{+\infty} \dd x\,Q_m(x) F(t-rx) \right]\nn\\ 
	& \qquad\quad - \frac{2(k-3-B)\hat{n}_L r^{B-k+1}}{(k+\ell-2-B)(k-\ell-3-B)} \bigg( \overset{(1)}{G}(t-r)\int_1^{+\infty} \dd x \, Q_m(x) F(t-rx) \nn\\
	& \qquad\qquad\qquad\qquad\qquad\qquad\qquad\qquad\quad
	+\ G(t-r)\int_1^{+\infty} \dd x \,x\, Q_m(x) \overset{(1)}{F}(t-rx) \bigg) \nn\\
	& \qquad\quad - \frac{\hat{n}_L r^{B-k+2}}{(k+\ell-2-B)(k-\ell-3-B)} \bigg(2\,\overset{(1)}{G}(t-r) \int_1^{+\infty} \dd x \,(x-1) Q_m(x) \overset{(1)}{F}(t-rx) \nn\\
	&\qquad\qquad\qquad\qquad\qquad\qquad \qquad 
	+   G(t-r)\int_1^{+\infty} \dd x \,(x^2-1)\, Q_m(x) \overset{(2)}{F}(t-rx) \bigg) \,.
\end{align}
This formula, when applied iteratively, allows us to reduce any case $k\geqslant 3$ in terms of the cases when $k=1$ and $2$, modulo a series of all-integrated terms (\emph{i.e.}, inside the d'Alembertian operator). Note that we introduced a new type of integral, $\int_1^{+\infty} \dd x \,x^n\, Q_m(x) H(t-rx)$ where $n \in \mathbb{N}$, but which can easily be recast in the previous form (\emph{i.e.} $n=0$) by recursively applying Bonnet's recursion formula for the Legendre function, namely
\begin{subequations}
\label{eq:Bonnet}
\begin{align}
	\label{subeq:BonnetLargerThan1}
	x \,Q_m(x) &= \frac{m+1}{2m+1}Q_{m+1}(x)+\frac{m}{2m+1}Q_{m-1}(x)& \text{if $m \geqslant 1$}\,, \\
	\label{subeq:Bonnet0}
	x \,Q_0(x) &= Q_1(x)+1 &  \text{if $m =0$}\,.
\end{align}
\end{subequations}

An important point is that we keep the Hadamard regulator $r^B$ ``alive''  in~\eqref{eq:kReducingIdentityToM}. Indeed, by applying the inverse d'Alembertian operator on both sides of~\eqref{eq:kReducingIdentityToM}, we see that the first term in the right side being the d'Alembertian of a source term containing the regulator $r^B$, will directly yield that source term proportional to $r^B$ without any additional homogeneous solution (since no homogeneous solution can be proportional to $r^B$). Hence the iteration with $r^B$ can be done blindly, ignoring homogeneous solutions, and only at the end do we apply the finite part when $B\to 0$. 

However a price we have to pay is that the $B$-dependent coefficients in~\eqref{eq:kReducingIdentityToM} can generate a simple pole when $B\to 0$ which will compete with the higher contribution $\propto B$ in the retarded integral, and \emph{vice versa}, a term $\propto B$ in the coefficient will be compensated by a pole coming from the retarded integral. A useful fact to remember in this respect is that the integrals are convergent for $k \in \{1,2\}$ and hence no poles $\propto 1/B$ can be generated in these cases. Furthermore, the structure of the identity \eqref{eq:kReducingIdentityToM} can only generate simple poles $\propto 1/B$, and no double poles $\propto 1/B^2$ or any poles of higher order. The presence of these simple poles is the main reason why we introduced the ${}_{k,m}\chi_L$ integral, as is clear by its first definition in~\eqref{eq:defChikmL}.

Finally, although we now have all the needed formulae to integrate the tails-of-memory, our recursion formula~\eqref{eq:kReducingIdentityToM} coupled to the $m=0$ case of the Bonnet formula~\eqref{subeq:Bonnet0} can generate instantaneous terms in the source, which must also be integrated. Although the integration of such terms is well known using standard integration techniques~\cite{BD92,B98quad}, its generalization to the case where simple poles $\propto 1/B$ can appear in the source was unknown. 

\subsection{Application to tails-of-tails}
\label{subsec:TailsOfTailsAndTailsTechniques}

The previous observation motivates us to extend our formalism developed for complicated tails-of-memory to the easier cases of tails-of-tails, tails and even instantaneous terms, such that we can treat the entire problem consistently, using one single formalism. When comparing with previous works~\cite{B98tail,MBF16} this will provide important tests of the results of Sec.~\ref{subsec:ToMTechniques}. Thus, we specialize the formulae of the previous section to the case where $G\equiv 1$, since in the case of the tails-of-tails two of the moments are just the mass $\dM$. 

When $k \geqslant 3$, we can again reduce the source to the cases $k=1$ and $k=2$ using \eqref{eq:kReducingIdentityToM}, with now $G^{(1)}(u) = 0$. Then, after lengthy computations where we perform two of the three integrations of the tails-of-memory (one on the  $y$ variable in the kernel and one on the time variable $\tau$), we find drastically simpler expressions when $k\in\{1,2\}$, which read 
\begin{subequations}
	\begin{align}
		\underset{2,m}{\ \overline{\Psi}_\ell}{\bigg|}_{G=1} &=2\, \underset{{2,m}}{\alpha_\ell} \overset{(-1)}{F}(u) \,, \\
		\underset{2,m}{\ \overline{\chi}_\ell}{\bigg|}_{G=1}  &= 2 \int_0^{+\infty}\dd \tau\, \left[\underset{{2,m}}{\beta_\ell} \ln\left(\frac{\tau}{2r_0}\right)    +  \underset{{2,m}}{\gamma_\ell} \right] F(u-\tau)  \,,\\
		\underset{1,m}{\ \overline{\Psi}_\ell}{\bigg|}_{G=1} &=\frac{1}{4} \int_0^{+\infty} \dd \tau \, \overset{(-1)}{F}(u-\tau)  \left[ \ln^2\left(\frac{\tau}{2r_0}\right)+4 H_m \ln\left(\frac{\tau}{2r_0}\right)-8\underset{{1,m}}{\delta_\ell} \right]\,,\\
		\underset{1,m}{\ \overline{\chi}_\ell}{\bigg|}_{G=1}  &=  \frac{1}{12}  \int_0^{+\infty} \dd \tau \, \overset{(-1)}{F}(u-\tau) \bigg[ \ln^3\left(\frac{\tau}{2r_0}\right) + 6 H_m \ln^2\left(\frac{\tau}{2r_0}\right) -24 \underset{{1,m}}{\delta_\ell}  \ln\left(\frac{\tau}{2r_0}\right)+12 \underset{{1,m}}{\varepsilon_\ell} \bigg]\,.
	\end{align}
\end{subequations}
where ${}_{k,m}{\overline{\Psi}_L} \equiv - \frac{\hat{n}_L}{2r}\ {}_{k,m}{\overline{\Psi}_\ell}$ and  ${}_{k,m}{\overline{\chi}_L} \equiv - \frac{\hat{n}_L}{2r}\ {}_{k,m}{\overline{\chi}_\ell}$.
The numerical constants ${}_{2,m}\alpha_\ell$, ${}_{2,m}\beta_\ell$ and ${}_{2,m}\gamma_\ell$ are defined by the integrals
\begin{subequations}
	\begin{align}
		\underset{2,m}{\ \alpha_\ell} & \equiv \underset{2,m}{\ \beta_\ell}  \equiv \int_1^{+\infty} \dd x\, Q_m(x)Q_\ell(x)\,,\\
		\underset{2,m}{\ \gamma_\ell} & \equiv \frac{1}{2^{\ell+1}} \int_1^{+\infty} \dd x\, Q_m(x) \int_{-1}^1 \dd z \,\frac{(1-z^2)^\ell}{(x-y)^{\ell+1}}\left( - \ln\left(\frac{x-y}{2}\right)+H_\ell \right) \,.
	\end{align}
\end{subequations}
Note that ${}_{2,m}\alpha_\ell$ is a special case of $\,_{k,m}\alpha_\ell$ defined for $2 \leqslant k \leqslant \ell+2$ in (A16) of~\cite{B98tail}, whereas $\,_{2,m}\beta_\ell$ and $\,_{2,m}\gamma_\ell$ are an extension of (A20) of~\cite{B98tail}, in which these constants were defined only for $k \geqslant \ell+3$. In the case $k=2$, we find it natural to define ${}_{2,m}\alpha_\ell$ and ${}_{2,m}\beta_\ell$ to be equal.
\newpage
Similarly one can define ${}_{1,m}\delta_\ell$ and ${}_{1,m}\varepsilon_\ell$ using formal integrals but at the price of introducing a Hadamard ``partie finie'' (pf) to cure the bound of the integrals at $x=1$: 
\begin{subequations}
	\begin{align}
		\underset{1,m}{\ \delta_\ell} &\equiv \mathrm{pf} \int_{1}^{+\infty} \dd x\, Q_m(x) \frac{\dd Q_\ell}{\dd x}\,,\\
		\underset{1,m}{\ \varepsilon_\ell} &\equiv \mathrm{pf} \int_{1}^{+\infty} \dd x\, Q_m(x) \frac{\dd^2 S_\ell}{\dd^2x}\,.
	\end{align}
\end{subequations}
The Hadamard partie finie is defined in the usual way by removing the divergent part of the integral (and using for our purpose here a Hadamard scale equal to 1):
\begin{subequations}
	\begin{align}
	\mathrm{pf}\int_{1}^{+\infty} \dd x\, Q_m(x) \frac{\dd Q_\ell}{\dd x} \equiv&	\lim_{\eta\to 0}\left\{\int_{1+\eta}^{+\infty} \dd x\, Q_m(x) \frac{\dd Q_\ell}{\dd x} +\frac{1}{8} \ln^2 \left(\frac{\eta}{2}\right)+\frac{1}{2} H_m \ln \left(\frac{\eta}{2}\right) \right\}\,, \\
	\mathrm{pf}\int_{1}^{+\infty} \dd x\, Q_m(x) \frac{\dd^2 S_\ell}{\dd^2x}\equiv& \lim_{\eta\to 0}\left\{ \int_{1+\eta}^{+\infty} \dd x\, Q_m(x) \frac{\dd^2 S_\ell}{\dd^2x} + \frac{1}{6}\ln^3 \left(\frac{\eta}{2}\right) + \frac{1}{2}H_m \ln^2 \left(\frac{\eta}{2}\right) \right\}\,.
	\end{align}
\end{subequations}
In the definition of ${}_{1,m}\varepsilon_\ell$ the function $S_\ell$ is defined like for the Neumann formula for the Legendre function [see \eqref{eq:Qm}], 
\begin{align}
	S_\ell(x) &\equiv \frac{1}{2}\int_{-1}^{1} \dd y\, P_\ell(y) \ln^2\left(\frac{x-y}{2}\right)\nn\\
	&=  \sum_{j=0}^\ell \sum_{i=0}^j \frac{(-)^{i+1}P_\ell^{(j)}(1)}{2(i+1)!(j-i)!}\Bigg\{(x-1)^{j+1}\left[\ln^2\left(\frac{x-1}{2}\right)- \frac{2}{i+1}\ln\left(\frac{x-1}{2}\right) + \frac{2}{(i+1)^2}\right]\nn\\
	&\qquad + (-)^{j+\ell+1}(x+1)^{j+1}\left[\ln^2\left(\frac{x+1}{2}\right)- \frac{2}{i+1}\ln\left(\frac{x+1}{2}\right) + \frac{2}{(i+1)^2}\right] \Bigg\}\,.
\end{align}
The constant ${}_{1,m}\delta_{\ell}$ was used in (A7-A9) of~\cite{B98tail} in the special case where $m=\ell$, but was not given a name. The constants defined in~\cite{B98tail} were later shown in~\cite{MBF16} to be sometimes ill-defined, \emph{e.g.} for $k=0$, $\ell=0$ and $m=0$, and this problem was circumvented on a case-by-case basis. Since we restrict our attention to $k\in \{1,2\}$, we are assured that our constants are always well defined.

All the coefficients we shall need can be computed analytically. The ${}_{2,m}\alpha_\ell $ constant (and, hence, $\,_{2,m}\beta_\ell $) has a closed form expression given by (A18) of~\cite{B98tail}, which reads
\begin{equation}
	\underset{2,m}{\ \alpha_\ell}  =  \underset{2,m}{\ \beta_\ell}  = \begin{cases}\displaystyle\frac{H_m - H_\ell}{(m-\ell)(m+\ell+1)} & \text{if } m\neq \ell\,, \\[0.3cm] \displaystyle\frac{1}{2m+1}\left(\frac{\pi^2}{6}-\sum_{j=1}^m \frac{1}{j^2}\right)& \text{if } m=\ell\,.\end{cases}
\end{equation}
The ${}_{1,m}\delta_\ell $ constant has a simple expression when $m=\ell$, as pointed out in Eqs.~(A7--A9) of~\cite{B98tail}, which reads ${}_{1,\ell}\delta_\ell = -\frac{1}{2}H_\ell^2$.
%
When $m \neq \ell$, by integrating by parts the integral expression of the ${}_{1,m}\delta_\ell$ constant, we find the simple property  
\begin{subequations}
\begin{equation}\label{eq:property}
	\underset{1,m}{\ \delta_\ell} + \underset{1,\ell}{\ \delta_m}  = - H_m H_\ell\,.
\end{equation}
We can then obtain the full expression of ${}_{1,m}\delta_\ell$ by using recursively the differential equation $Q'_{n+1} - Q'_{n-1} = (2n+1) Q_n$. In the case where $\ell$ and $m$ have the same parity, \emph{i.e.} $\ell-m \in 2\mathbb{Z}$, the differential equation allows us to express the constant only in terms of the case ${}_{1,\ell}\delta_\ell$, along with many ${}_{2,m}\alpha_\ell$ terms, which are easy to compute. Thanks to \eqref{eq:property}, we can further restrict attention to the case where $\ell \geqslant m$, and the relevant formula then reads
\begin{equation} 
\underset{1,m}{\ \delta_\ell}  = \sum_{j=1}^{\frac{\ell-m}{2}} \bigl(2m +4j-1\bigr) \underset{2,m}{\ \alpha_{m+2j-1}}  - \frac{1}{2} H_m^2 \, ,
\end{equation}
if $\ell-m \in 2 \mathbb{N}$. In the reverse case where $\ell$ and $m$ have opposite parity, \emph{i.e.} $\ell-m \in 2 \mathbb{Z} +1$, we can again use \eqref{eq:property} to restrict attention to the case where $\ell$ is odd and $m$ is even. In this case, we can again use the differential equation as well as integration by parts to express our constant solely in terms of the explicitly known case, ${}_{1,0}\delta_1 = \frac{\pi^2}{12}$, modulo some ${}_{2,m}\alpha_\ell$  terms. We thus find the relevant formula,
\begin{align}
\underset{1,m}{\ \delta_\ell}  = - H_m + \frac{\pi^2}{12} + \sum_{j = 1}^{\frac{\ell-1}{2}}\bigl( 2\ell -4j+3 \bigr)\underset{2,m}{\ \alpha}{}_{\ell-2j+1} - \sum_{j=1}^{\frac{m}{2}}\bigl(2m-4j+3\bigr)&\underset{2,1}{\ \alpha}{}_{m-2j+1}\,,
\end{align}
\end{subequations}
if $\ell \in 2 \mathbb{N} +1 $ and $m \in 2 \mathbb{N}$. Finally the ${}_{2,m}\gamma_\ell$ and ${}_{1,m}\varepsilon_\ell$ constants were only needed for the following values of~ $m$~and~$\ell$, which we calculated using \emph{Mathematica}:
\begin{align}
	\underset{2,0}{\ \gamma_1} =  \frac{5}{4}\,,&& \underset{2,2}{\ \gamma_1} = \frac{13}{32} \,, && \underset{2,4}{\ \gamma_1} = \frac{95}{432}\,,\nn  \\
	\underset{1,0}{\ \varepsilon_0} = \frac{\zeta(3)}{2}\,, && \underset{1,2}{\ \varepsilon_0} = \frac{9}{2}+\frac{\zeta(3)}{2}\,,  && \underset{1,4}{\ \varepsilon_0} = \frac{3995}{432}+ \frac{\zeta(3)}{2}\,,\nn \\
	\underset{1,1}{\ \varepsilon_1} = \frac{1}{2}+ \frac{\zeta(3)}{2}\,, && \underset{1,3}{\ \varepsilon_1} = \frac{299}{48}+ \frac{\zeta(3)}{2}\,, && \underset{1,5}{\ \varepsilon_1}= \frac{887}{80}+ \frac{\zeta(3)}{2}\,,\nn \\
	\underset{1,0}{\ \varepsilon_2} = -\frac{15}{2}+ \frac{\zeta(3)}{2}\,, && \underset{1,2}{\ \varepsilon_2} = \frac{33}{16}+ \frac{\zeta(3)}{2}\,, && \underset{1,4}{\ \varepsilon_2} =  \frac{3425}{432}+ \frac{\zeta(3)}{2}\,,
\end{align}
where $\zeta$ is the Riemann function and $\zeta(3)$ is the Ap\'ery constant. In our computation, the contributions proportional to this constant actually cancel out.

We end this section with a word concerning the integration of $B$-regularized source terms that are actually instantaneous, \emph{i.e.} just of the simple type ${}_{k}S^B(r,t-r) \equiv (\frac{r}{r_0})^B \,r^{-k} F(t-r)$ (multiplied by the multipolarity factor $\hat{n}_L$). These source terms yield for instance the tails at quadratic order. In line with our general formalism, if $k \geqslant 3$, we bring ourselves to the case $k=2$ using the recursion formula:
\begin{align}
\label{eq:kReducingIdentityTails}
	\hat{n}_L r^{B-k} F(t-r) &= \Box\left[ \frac{\hat{n}_L r^{B-k+2}}{(k+\ell-2-B)(k-\ell-3-B)} F(t-r) \right] \nn\\
	&\quad\ - \frac{2(k-3-B)\hat{n}_L r^{B-k+1}}{(k+\ell-2-B)(k-\ell-3-B)}\overset{(1)}{F}(t-r)\,. 
\end{align}
The case $k=2$ is well known \cite{BD92,B98quad,B98tail}, and for the asymptotic limit reads 
\begin{equation}
\label{eq:integrationFormulaTails}
\FPprop \biggl[\left(\frac{r}{r_0} \right)^B \frac{\hat{n}_L}{r^2}F(t-r) \biggr] = \frac{\hat{n}_L}{2r}\int_{0}^{+\infty}\dd \tau\, F(u-\tau)\Bigl[\ln\left(\frac{\tau}{2r}\right) + 2 H_\ell \Bigr] + o\left( \frac{1}{r}\right)\,,
\end{equation}
where we could have dropped the finite part prescription and set $B=0$, because in this case the inverse d'Alembertian integral is convergent. However, because of the potential appearance of single poles due to our recursion formula \eqref{eq:kReducingIdentityTails}, we will also need the corresponding formula with an extra factor $B^{-1}$ multiplying the source term. The relevant formula reads
\begin{align} 
	\label{eq:integrationFormulaTailsPole}
	&  \FPprop \biggl[\frac{1}{B}\left(\frac{r}{r_0} \right)^B \frac{\hat{n}_L}{r^2}F(t-r) \biggr] \\
	& \quad = \frac{\hat{n}_L}{4r} \int_{0}^{+\infty}\!\dd \tau\, F(u-\tau)\biggl[ \ln^2 \left(\frac{\tau}{2r_0}\right) + 4 H_\ell \ln \left(\frac{\tau}{2r_0}\right)+4 H_\ell^2 - \ln^2 \left( \frac{r}{r_0}\right) \biggr] + o\left( \frac{1}{r}\right)\,,\nn
\end{align}
which can be obtained from (A2) of \cite{B98quad} by taking straightforwardly the finite part (since the integral converges) and expanding when $r \rightarrow +\infty $.

\section{Implementing the calculation of tails-of-memory}
\label{sec:radiativeQuadrupole}

\subsection{Explicit integration of the asymptotic kernels}
\label{subsec:integrationAsymptoticKernels}

Up to this point, the kernels in the asymptotic limit $r\to+\infty$ were defined only in an integral form, given by~\eqref{eq:defAsympKernelsKbarLbar} in the relatively easy case where $k=2$, and by the more complex forms~\eqref{subeq:expressionK1bar} and~\eqref{subeq:expressionL1bar} when $k=1$. These integrals are too complicated to seek a general explicit formula valid for arbitrary $m$ and $\ell$. However, we can easily compute all these integrals on a case-by-case basis. For this it suffices to insert into them the explicit expressions of the Legendre polynomial $P_m(x)$ and the Legendre function of the second kind $Q_m(x)$, which are simply polynomials multiplied by some logarithms, see~\eqref{eq:Qm}. 

In this way we find that the general structure of the kernels ${}_{k,m}K_{\ell}(\rho,\tau,r)$ and ${}_{k,m}L_{\ell}(\rho,\tau,r)$ for $k=1$ and $2$, up to $o(1)$ precision, is of the type
\begin{equation}
	\left(\begin{array}{l} \displaystyle \underset{k,m}{\ K_{\ell}}(\rho,\tau,r) \\[0.7cm] \displaystyle \underset{k,m}{\ L_{\ell}}(\rho,\tau,r) \end{array}\right) = \sum_{s, p, q} \left(\begin{array}{l} \displaystyle \underset{k,m}{\ \mathcal{X}}\!\!{}_{\,s,p,q}\\[0.7cm] \displaystyle  \underset{k,m}{\ \mathcal{Y}}\!\!{}_{\,s,p,q}\end{array}\right)\int_{\frac{2\tau}{\rho+2r}}^{\frac{2\tau}{\rho}}\dd y\, y^s \ln^p \left(\frac{y}{2}\right)\ln^q \left(1+\frac{y}{2}\right) \,,
\end{equation}
where ${}_{k,m}\mathcal{X}_{s,p,q}$ and ${}_{k,m}\mathcal{Y}_{s,p,q}$ denote some numerical coefficients, with $s$ an integer such that $s\geqslant k-2$, and where we can restrict our attention only to the values $p=0,1,2,3$ and $q=0,1$. When $s\geqslant 0$, the lower bound of the integral can be set to $0$ in the $r\to+\infty$ limit, but not when $s = -1$ which can happen when $k=1$. In the latter case, the integral will develop the logarithmic behavior when $r\to+\infty$ which has already been obtained in~\eqref{subeq:expressionKernelK1Asymp} and~\eqref{subeq:expressionKernelL1Asymp}, and we now only consider the finite part. 

When $q=0$, these integrals are of course well-known (see \emph{e.g.} 2.721-722 in~\cite{GR}), so we now restrict our discussion to the case $q=1$. Let us first examine the case $q=1$ and $s\geqslant 0$; we thus set the lower bound to zero, in the limit $r\to+\infty$. We perform an integration by parts using 2.722 in~\cite{GR},
\begin{align}
	\int_{0}^{\frac{2\tau}{\rho}}&\dd y\, y^s  \ln^p \left(\frac{y}{2}\right) \ln \left(1+\frac{y}{2}\right) \\
	&= \sum_{i=0}^p \frac{(-)^i p!}{(p-i)!(s+1)^{i+1}}\Bigg\{ \left(\frac{2\tau}{\rho}\right)^{s+1}  \!\ln^{p-i} \left(\frac{\tau}{\rho}\right) \ln \left(1+\frac{\tau}{\rho}\right)- \int_{0}^{\frac{2\tau}{\rho}}\dd y\, \frac{y^{s+1}}{y+2}  \ln^{p-i} \left(\frac{y}{2}\right) \Bigg\}\,.\nn
\end{align}
The rational fraction $y^{s+1} (y+2)^{-1}$ in the remaining integral can be expanded as a polynomial, plus a function proportional to $(y+2)^{-1}$. From this result we see that the only non trivial integrals left are of the type $\int_{0}^{z}\dd  y\, (y+2)^{-1}  \ln^{j} \left(y/2\right)$, and these are simply related to the polylogarithm functions, for instance\footnote{The polylogarithm, or Jonqui\`ere's function, is defined for any $n \in \mathbb{N}$ and $z>0$ as
	$$\mathrm{Li}_n(z) \equiv \frac{(-)^{n-1}}{(n-2)!}\int_0^z \frac{ \dd s}{s} \, \ln^{n-2}\left(\frac{s}{z}\right)\ln(1-s)\,.$$ In particular, the dilogarithm, or Spence's function, is defined as $\mathrm{Li}_2(z) = - \int_0^z \frac{\dd s}{s}\, \ln(1-s)$. We also find that $\mathrm{Li}_1(z) = - \ln(1-z)$.}
\begin{align}
\int_{0}^{\frac{2\tau}{\rho}}\frac{\dd y}{y+2} \ln^2 \left(\frac{y}{2}\right) &= - 2 \mathrm{Li}_3\left(-\frac{\tau}{\rho}\right) + 2 \mathrm{Li}_2\left(-\frac{\tau}{\rho}\right)  \ln\left(\frac{\tau}{\rho}\right) - \mathrm{Li}_1\left(-\frac{\tau}{\rho}\right)\ln^2\left(\frac{\tau}{\rho}\right)\,.
\end{align}
Now, the only remaining cases are when $q=1$ and $s=-1$, but in these cases too the integrals are related to polylogarithms, for instance
\begin{align}
	\int_{0}^{\frac{2\tau}{\rho}}\frac{\dd y}{y}  \ln \left(\frac{y}{2}\right) \ln \left(1+\frac{y}{2}\right) = \mathrm{Li}_3\left(-\frac{\tau}{\rho}\right) - \,\mathrm{Li}_2\left(-\frac{\tau}{\rho}\right) \ln\left(\frac{\tau}{\rho}\right)\,.
\end{align}

To summarize we have obtained the following general structure of the kernel functions in the asymptotic limit $r\to+\infty$, discarding the leading logarithmic behavior in the $k=1$ case --- recall~\eqref{subeq:expressionKernelK1Asymp} and~\eqref{subeq:expressionKernelL1Asymp} ---, as
\begin{equation}\label{eq:structKbarre}
	\left(\begin{array}{l} \displaystyle \underset{k,m}{\ \overline{K_{\ell}}}(\rho,\tau) \\[0.7cm] \displaystyle \underset{k,m}{\ \overline{L_{\ell}}}(\rho,\tau) \end{array}\right) = \tau^{1-k}\sum_{\scriptsize{\begin{matrix} j \in \mathbb{Z} \\ p,q,n \in \mathbb{N}\end{matrix}} } \left(\begin{array}{l} \displaystyle \underset{k,m}{\ \kappa}{}_{j,p,q,n}\\[0.7cm] \displaystyle  \underset{k,m}{\ \lambda}{}_{j,p,q,n}\end{array}\right) \left(\frac{\tau}{\rho}\right)^j \ln^p\left(\frac{\tau}{2 r_0}\right) \ln^q\left(\frac{\rho}{2 r_0}\right)\, \overline{\mathrm{Li}}_n\left(-\frac{\tau}{\rho}\right)\,,
\end{equation}
with some numerical coefficients ${}_{k,m}\kappa_{j,p,q,n}$ and ${}_{k,m}\lambda_{j,p,q,n}$ that depend on the relative integer $j\in\mathbb{Z}$ and natural integers $p,q,n\in\mathbb{N}$, and where we have defined for notational convenience \mbox{$\overline{\mathrm{Li}}_0(z)\equiv 1$} and \mbox{$\overline{\mathrm{Li}}_n(z)\equiv\mathrm{Li}_n(z)$} for $n\geqslant 1$. We also recall that $\mathrm{Li}_1(z) = -\ln(1-z)$.

The coefficients ${}_{k,m}\kappa_{j,p,q,n}$ and ${}_{k,m}\lambda_{j,p,q,n}$ represent six-dimensional matrices of numerical coefficients and cannot be presented. Instead, we provide the explicit expressions of all the kernels needed for the computation of the tails-of-memory in the Supplementary Material \cite{SuppMaterial}. For the purpose of clarity, we give two explicit examples of such kernels, where we choose to look at $\overline{K}$ for $m=2$, $\ell=4$ and $k = 1,2$: 
\begin{subequations}
\begin{align}
\underset{1,2}{\ \overline{K_{4}}}(\rho,\tau)  =& - \frac{1}{2} \,\mathrm{Li}_2\left(-\frac{\tau}{\rho}\right)+ \frac{7}{12} \,\mathrm{Li}_1\left(-\frac{\tau}{\rho}\right) - \frac{1}{4} \ln^2\left(\frac{\rho}{2 r_0}\right) + \frac{1}{2} \ln\left(\frac{\rho}{2 r_0}\right) \ln\left(\frac{\tau}{2 r_0}\right) \nn\\
& - \frac{7}{12}  \ln\left(\frac{\rho}{2 r_0}\right) + \frac{25}{12} \ln\left(\frac{\tau}{2 r_0}\right) + \frac{9}{4} - \frac{7}{12}\,  \frac{\rho}{\tau}+\frac{35}{24} \left(\frac{\rho}{\tau}\right)^2 \nn \\
&+\left(\frac{\rho}{\tau}\right)^3\left[\frac{7}{3}\,\mathrm{Li}_1\left(-\frac{\tau}{\rho}\right) + \frac{7}{4}\right] +\frac{7}{4}\,\left(\frac{\rho}{\tau}\right)^4 \mathrm{Li}_1\left(-\frac{\tau}{\rho}\right) \,, \\
\underset{2,2}{\ \overline{K_{4}}}(\rho,\tau)  =&\ \frac{1}{\tau}\Bigg\{\frac{1}{6} - \frac{5}{6}\, \frac{\rho}{\tau} + \left(\frac{\rho}{\tau}\right)^2 \left[-3\,\mathrm{Li}_1\left(-\frac{\tau}{\rho}\right)-5\right]\nn \\
&\qquad+ \left(\frac{\rho}{\tau}\right)^3 \left[-7\,\mathrm{Li}_1\left(-\frac{\tau}{\rho}\right)-4\right] - 4\left(\frac{\rho}{\tau}\right)^4 \,\mathrm{Li}_1\left(-\frac{\tau}{\rho}\right) \Bigg\}\,.
\end{align}
\end{subequations}
Notice that despite the structure in $\rho/\tau$ of these kernels, one can check explicitly that they are indeed integrable in a vicinity of $(\rho,\tau)=(0,0)$.

\subsection{Raw expression of the radiative quadrupole in the radiative construction}
\label{subsec:rawRadiativeResults}

We have applied the MPM construction of the metric in radiative gauge as described in Sec.~\ref{sec:radMPM}, together with the previous integration techniques, to the computation of the three cubic interactions $\dM^2 \times \dMbar_{ij}$, $\dM\times\dS_i\times \dMbar_{ij}$ and $\dM\times\dMbar_{ij}\times \dMbar_{ij}$.

The first interaction, $\dM^2 \times \dMbar_{ij}$, is the tail-of-tail which enters at 3PN order and was already known in the harmonic construction~\cite{B98tail}. We have, on the one hand, computed the asymptotic waveform for this interaction in the radiative construction using the integration machinery developed in Sec.~\ref{sec:techniques}, and, on the other hand, used standard integration techniques~\cite{B98quad, B98tail, FBI15, MBF16} applied to the radiative algorithm, and checked that both methods yielded 
identical results. We then verified that these results, performed in the radiative construction, could be independently retrieved from the known result in the harmonic construction~\cite{B98tail} solely using a moment redefinition, as explained in Ref.~\cite{TLB22}. In the radiative gauge, we find
\begin{equation}\label{subeq:UijRadMxMxMbarij}
	\mathcal{U}_{ij}^{\dM^2 \times \dMbar_{ij}} = \frac{2 G^2\dM^2}{c^6} \int_{0}^{+\infty} \!\! \dd\tau \left[\ln^2\left(\frac{\tau}{2r_0}\right) + \frac{107}{42}\ln\left(\frac{\tau}{2r_0}\right) + \frac{40037}{8810}\right]\dMbar_{ij}^{(5)}(u-\tau)\,.
\end{equation}

For the $\dM\times\dS_i\times \dMbar_{ij}$ interaction, which only enters at 4PN, the fact that the current dipole moment $\dS_i$ (or total angular momentum) is constant greatly simplifies the computation, and \emph{a priori} allows for the use of standard techniques~\cite{B98quad, B98tail, MBF16}. However, in some cases, those techniques break down, since some of the numerical constants introduced are ill-defined, as has been mentioned in~\cite{MBF16}. Therefore, we have computed for the first time the full  $\dM\times\dS_i\times \dMbar_{ij}$ waveform using our new integration method, which reads in radiative gauge,
\begin{equation}\label{subeq:UijRadMxSixMbarij}
	\mathcal{U}_{ij}^{\dM\times\dS_i \times \dMbar_{ij}} = - \frac{2 G^2 \dM}{c^8} \,\dS_{a\vert \langle i} \int_{0}^{+\infty} \dd\tau\, \left[ \ln\left(\frac{\tau}{2r_0}\right) + \frac{5381}{5670} \right] \dMbar_{j\rangle a}^{(6)}(u-\tau)\,.
\end{equation}

Finally, the $\dM\times\dMbar_{ij}\times \dMbar_{ij}$ interaction is the genuine tail-of-memory entering at 4PN; the double time-dependence within the two associated quadrupole moments strongly complicates the situation. We have performed this calculation only in the asymptotic limit $r\to+\infty$. Since we are using the radiative construction~\cite{B87} the waveform is automatically free of far-zone logarithms, and can straightforwardly be projected in a TT gauge, where the radiative quadrupole moment $\mathcal{U}_{ij}$ is extracted in a standard way. Here we first present the ``raw'' result, obtained in a direct manner from our integration formulae, in terms of the radiative canonical moments. In Sec.~\ref{subsec:simplifcationMethod} we describe a very efficient simplification method, and the resulting radiative quadrupole will be presented in Sec.~\ref{sec:results} in terms of the harmonic canonical moments. The ``raw'' result is however important and worthy to be presented, as it is the direct result of our integration scheme. 

In order to present the raw result, we define the following functionals of two time-derivatives of quadrupole moments $F$ and $G$ for the cases $k=1,2$:
\begin{subequations}\label{eq:defPsiChibar}
	\begin{align}
	\label{subeq:defPsibar}
		\underset{k,m}{\ \overline{\Psi}_\ell}[F,G] &\equiv \int_{0}^{+\infty} \dd\rho\,G(u-\rho)\int_{0}^{+\infty}   \dd\tau\,F(u-\rho-\tau) \underset{k,m}{\ \overline{K}_{\ell}}(\rho,\tau)\,,\\
	\label{subeq:defChibar}
		\underset{k,m}{\ \overline{\chi}_\ell}[F,G] &\equiv \int_{0}^{+\infty} \dd\rho\,G(u-\rho)\int_{0}^{+\infty}   \dd\tau\,F(u-\rho-\tau) \underset{k,m}{\ \overline{L}_{\ell}}(\rho,\tau)\,,
	\end{align}
\end{subequations}
where the kernels are given by~\eqref{eq:defAsympKernelsKbarLbar} in the case $k=2$ and by Eqs.~\eqref{subeq:expressionK1bar} and~\eqref{subeq:expressionL1bar} in the case $k=1$. As explained in Sec.~\ref{subsubsec:ToMTechniquesCasek3} we do not need the cases $k\geqslant 3$.

With these definitions in hand, we decompose the cubic interaction $\dM\times\dMbar_{ij}\times \dMbar_{ij}$ as follows: we separate terms which are purely instantaneous (\emph{i.e.} local-in-time), from terms whose non-locality is rather simple and looks like that for the ordinary tails or tails-of-tail, and finally from the genuine and much more intricated tail-of-memory (ToM) integrals. We thus write our raw result, expressed in terms of the radiative canonical moments, as
\begin{equation}\label{subeq:UijRadMxMbarijxMbarij}
	\mathcal{U}_{ij}^{\dM\times\dMbar_{ij} \times \dMbar_{ij}} = \mathcal{U}_{ij}^{\dM\times\dMbar_{ij} \times \dMbar_{ij}}{\bigg|}_\text{inst} + \mathcal{U}_{ij}^{\dM\times\dMbar_{ij} \times \dMbar_{ij}}{\bigg|}_\text{tail} + \mathcal{U}_{ij}^{\dM\times\dMbar_{ij} \times \dMbar_{ij}}{\bigg|}_\text{ToM} \,.
\end{equation}
\begin{itemize}
\item[1.] The instantaneous terms (depending only on the current time $u$) are given by
\begin{subequations}
	\begin{align}
		\label{subeq:UijRadMxMbarijxMbarijInst}
		\mathcal{U}_{ij}^{\dM\times\dMbar_{ij} \times \dMbar_{ij}}{\bigg|}_\text{inst}  &= \frac{G^2\dM}{c^8}\biggl[\frac{3362032}{165375} \dMbar_{a \langle i}^{(3)}  \dMbar_{j \rangle a}^{(3)} + \frac{2014871}{165375}\dMbar_{a \langle i}^{(2)}  \dMbar_{j \rangle a}^{(4)}   \nn\\&\qquad\quad -  \frac{5766241}{165375}\dMbar_{a \langle i}^{(1)}  \dMbar_{j \rangle a}^{(5)}   - \frac{114454}{23625} \dMbar_{a \langle i}  \dMbar_{j \rangle a}^{(6)}\biggr]\,.
\end{align}
\item[2.] The non-local tail like terms involve some logarithmic kernels, and naturally the quadrupole moment of the usual tail terms is here replaced by a combination $\propto \dMbar_{a\langle i}^{(n)} \,\dMbar_{j \rangle a}^{(p)}$ evaluated at any time $u-\tau$ in the past. We have
\begin{align}
		\label{subeq:UijRadMxMbarijxMbarijTail}
		\mathcal{U}_{ij}^{\dM\times\dMbar_{ij} \times \dMbar_{ij}}{\bigg|}_\text{tail}  =& - \frac{48}{175} \frac{G^2\dM}{c^8} \int_{0}^{+\infty} \dd \tau \left[\ln^2\left(\frac{\tau}{2 r_0}\right) + \frac{5111}{840}\ln \left(\frac{\tau}{2 r_0}\right)  \right] (\dMbar_{a\langle i} \,\dMbar_{j \rangle a}^{(7)})(u-\tau)\nn\\
		& - \frac{192}{175}  \frac{G^2\dM}{c^8} \int_{0}^{+\infty} \dd \tau \left[\ln^2\left(\frac{\tau}{2 r_0}\right) + \frac{475861}{30240}\ln \left(\frac{\tau}{2 r_0}\right)  \right] (\dMbar_{a\langle i}^{(1)} \,\dMbar_{j \rangle a}^{(6)})(u-\tau)\nn\\
		& - \frac{3154}{315}  \frac{G^2\dM}{c^8} \int_0^{+\infty} \dd \tau \ln\left(\frac{\tau}{2 r_0}\right) (\dMbar_{a \langle i}^{(2)} \,\dMbar_{j \rangle a}^{(5)})(u-\tau) \nn\\
		& + \frac{478}{35} \frac{G^2\dM}{c^8} \int_0^{+\infty} \dd \tau \ln\left(\frac{\tau}{2 r_0}\right) (\dMbar_{a \langle i}^{(3)} \,\dMbar_{j \rangle a}^{(4)}) (u-\tau) \,.
\end{align}
\item[3.] The genuine tail-of-memory part is fully specified by the bilinear functionals of the two quadrupole moments defined by~\eqref{eq:defPsiChibar}. Thus we can write 
\begin{align}
	\label{subeq:UijHarmMxMijxMijToM}
	\mathcal{U}_{ij}^{\dM\times\dMbar_{ij} \times \dMbar_{ij}}{\bigg|}_\text{ToM}  &=  \frac{G^2\dM}{c^8}\sum_{m,\ell,n} \bigg\{ \mathcal{A}_{m,\ell}^{n}\,  \underset{1,m}{\ \overline{\Psi}_\ell} \bigl[\dMbar_{a \langle i}^{(n)},\dMbar_{j \rangle a}^{(8-n)}\bigr] + \mathcal{B}_{m,\ell}^{n}\,  \underset{2,m}{\ \overline{\Psi}_\ell} \bigl[\dMbar_{a \langle i}^{(n)},\dMbar_{j \rangle a}^{(7-n)}\bigr] \nn\\
	&\qquad\quad + \mathcal{C}_{n,\ell}^{n}\,  \underset{1,m}{\ \overline{\chi}_\ell}  \bigl[\dMbar_{a \langle i}^{(n)},\dMbar_{j \rangle a}^{(8-n)}\bigr]  + \mathcal{D}_{m,\ell}^{n}\,  \underset{2,m}{\ \overline{\chi}_\ell}  \bigl[\dMbar_{a \langle i}^{(n)},\dMbar_{j \rangle a}^{(7-n)}\bigr] \bigg\}\, ,
\end{align}
\end{subequations}
where the purely numerical coefficients $\mathcal{A}_{m,\ell}^{n}$, $\mathcal{B}_{m,\ell}^{n}$, $\mathcal{C}_{m,\ell}^{n}$ and $\mathcal{D}_{m,\ell}^{n}$ in front of each of these integrals are provided in the Tables~\ref{table:coeffs}, and the functionals ${}_{k,m}\overline{\Psi}_\ell$ and ${}_{k,m}\overline{\chi}_\ell$ are defined in terms of the kernels in \eqref{eq:defPsiChibar}. Thus our complete results follow from these Tables together with the explicit expressions of the kernel functions provided in the Supplementary Material \cite{SuppMaterial}.
\end{itemize}
\newpage
\subsection{Simplification method}
\label{subsec:simplifcationMethod}

In this Section, we implement a method for simplifying the expression of the pure \emph{tail-of-memory} part of the radiative quadrupole, given by~\eqref{subeq:UijHarmMxMijxMijToM}. The idea is to rexpress everything as only one double integral over the two quadrupoles and a single kernel, modulo some easy surface terms. We thus alternatively integrate by parts the $\rho$ and $\tau$ variables of~\eqref{eq:defPsiChibar}, so as to transfer all the time derivatives on the quadrupole moment represented by $F(u-\rho-\tau)$, \emph{i.e.} the left slot in the functionals~\eqref{eq:defPsiChibar}.

First we observe in~\eqref{subeq:UijHarmMxMijxMijToM} that when  $k=1$, the two quadrupoles have respectively $n$ and $8-n$ time derivatives, while in the case $k=2$, they instead have $n$ and $7-n$ derivatives. We first uniformize this by transforming the case $k=2$ with the formula
\begin{align}
\label{eq:IBPfromk2tok1}
\int_0^{+\infty} \dd \rho\, &\dMbar_{a\langle i}^{(n)}(u-\rho) \int_0^{+\infty} \dd \tau\, \dMbar_{j \rangle a}^{(7-n)}(u-\rho-\tau)\, f(\rho,\tau) \nn \\
& = \int_0^{+\infty} \dd \rho\, \dMbar_{a\langle i}^{(n)}(u-\rho) \int_0^{+\infty} \dd \tau\, \dMbar_{j \rangle a}^{(8-n)}(u-\rho-\tau)\, \bigl(\partial_\tau^{-1} f\bigr)(\rho,\tau)\,,
\end{align}
where we have introduced the $\tau$-antiderivative which vanishes at $\tau=0$, defined for any function $f(\rho,\tau)$ with adequate regularity properties as
\begin{equation}
\label{eq:tauAntiderivative}
\bigl(\partial_\tau^{-1} f\bigr)(\rho,\tau) \equiv \int_0^\tau \dd \lambda \,f(\rho, \lambda)\,. 
\end{equation}

After performing this operation, we are left with integrals of the moments that only have $n$ and $8-n$ time derivatives, and the only possible cases are $n\in [\![0,4]\!]$. Next we integrate by parts so as to be left with integrals with only $0$ and $8$ time derivatives. However this operation yields some integrals over $\rho$ that are separately divergent at the lower bound $\rho=0$. To cure this, we introduce a regularization and replace the $0$ at the lower bound by some small $\epsilon$, and restrict attention to the $\epsilon \rightarrow 0$ expansion. In the end, we will verify that the final result has a finite $\epsilon \rightarrow 0$ limit. With this caveat in mind, we will simplify all our integrals with the formula (valid for $n\in [\![0,4]\!]$)
\begin{align}
\label{eq:IBPmasterFormula}
&\int_\epsilon^{+\infty} \dd \rho\, \dMbar_{a\langle i}^{(n)}(u-\rho) \int_0^{+\infty} \dd \tau\, \dMbar_{j\rangle a}^{(8-n)}(u-\rho-\tau)\, f(\rho,\tau) \nn\\
&\qquad = \int_\epsilon^{+\infty} \dd \rho\, \dMbar_{a\langle i}(u-\rho) \int_0^{+\infty} \dd \tau\, \dMbar_{j\rangle a}^{(8)}(u-\rho-\tau)\, \mathcal{O}_{\rho,\tau}^{\,n} \bigl[f(\rho,\tau) \bigr]\nn\\
&\qquad \quad+ \sum_{k=0}^{n-1}  \dMbar_{a\langle i}^{(k)}(u-\epsilon) \int_0^{+\infty} \dd \tau \, \dMbar_{j\rangle a}^{(7-k)}(u-\tau-\epsilon) \bigg[\mathcal{O}_{\rho,\tau}^{\,n-1-k} \bigl[f(\rho,\tau) \bigr] \bigg]_{\rho=\epsilon}\,,
\end{align}
where moments involving $\epsilon$ should be Taylor-expanded when $\epsilon \rightarrow 0$, the second term is evaluated at $\rho=\epsilon$, and we have introduced the differential operator $\mathcal{O}_{\rho,\tau}$ defined by 
\begin{equation}
\label{eq:OdiffOperator}
\mathcal{O}_{\rho,\tau} \bigl[f(\tau,\rho) \bigr]\equiv \partial_\tau^{-1} \partial_\rho  \bigl[f(\tau,\rho)\bigr] - f(\tau,\rho)\,,
\end{equation}
together with its iterations $\mathcal{O}^{\,n}_{\rho,\tau}\equiv\mathcal{O}_{\rho,\tau}\cdots\mathcal{O}_{\rho,\tau}$. Applying the formula~\eqref{eq:IBPmasterFormula} to all the terms composing the tails-of-memory~\eqref{subeq:UijHarmMxMijxMijToM}, we arrive at a unique master double integral:
\begin{align}
\label{eq:UijHarmMxMijxMijToMv2}
\mathcal{U}_{ij}^{\dM\times\dMbar_{ij} \times \dMbar_{ij}}{\bigg|}_\text{ToM} = \dM \int_\epsilon^{+\infty} \dd \rho\, \dMbar_{a \langle i}(u-\rho) \int_0^{+\infty} \dd \tau\, \dMbar_{j\rangle a}^{(8)}(u-\rho-\tau)\, \Omega(\rho,\tau) + \mathcal{S}_\epsilon\,,
\end{align}
where $\Omega(\rho,\tau)$ denotes some new kernel function and $\mathcal{S}_\epsilon$ are all the surface terms coming from the second line of~\eqref{eq:IBPmasterFormula}, which carry at most one integral and that simplify drastically in the $\epsilon \rightarrow 0$ expansion. Since the expression is long we do not show here the result for $\mathcal{S}_\epsilon$.

At this stage, we would \emph{a priori} expect $\Omega(\rho,\tau)$ to have a very complicated structure akin to~\eqref{eq:structKbarre}, and in particular to involve many polylogarithms. But instead, we find the following simple expression without any polylogarithms:
\begin{align}
\label{eq:OmegaKernel}
\Omega(\rho,\tau) &= \frac{7 613 764}{165 375}-\frac{1 024 076}{18 375}\, \frac{\tau}{\rho}-\frac{2074}{63}\left(\frac{\tau}{\rho}\right)^2 -\frac{104}{15} \left(\frac{\tau}{\rho}\right)^3\nn\\
& + \frac{634 076}{55125}\ln\left(\frac{\rho}{2r_0}\right)+ \frac{384}{175}\, \frac{\tau}{\rho} \ln\left(\frac{\rho}{2r_0}\right)- \frac{144}{175} \ln\left(\frac{\rho}{2r_0}\right)^2+\frac{8}{7} \ln\left(\frac{\tau}{2r_0}\right)\,.
\end{align}
Note that if we changed even a single coefficient in Table~\ref{table:coeffs}, the cancellation of polylogarithms would not occur in general, and we would be left with a much more complicated expression for $\Omega(\rho,\tau)$. 
When inserting $\Omega(\rho,\tau)$ into the first term in~\eqref{eq:UijHarmMxMijxMijToMv2}, we can integrate by parts so as to remove all the powers of $\tau/\rho$. This introduces poles in $\epsilon$, and powers of the logarithms of $\epsilon$, but we have checked that these poles and divergences exactly cancel when adding the surface terms $\mathcal{S}_\epsilon$, namely the second term in~\eqref{eq:UijHarmMxMijxMijToMv2}. Putting all of this together and taking the $\epsilon \rightarrow 0$ limit, we find that
\begin{align}
\label{eq:UijHarmMxMijxMijToMv3}
\mathcal{U}_{ij}^{\dM\times\dMbar_{ij} \times \dMbar_{ij}}{\bigg|}_\text{ToM} &= \frac{8}{7}\,\dM \int_0^{+\infty} \dd\rho\,  \dMbar_{a \langle i}^{(4)}(u-\rho) \int_0^{+\infty} \dd \tau\, \ln\left(\frac{\tau}{2 r_0}\right) \dMbar_{j \rangle a}^{(4)}(u-\rho-\tau) \nn\\
&+\frac{48}{175}\,\dM \int_0^{+\infty} \dd\tau \,\dMbar_{a \langle i} \dMbar^{(7)}_{j \rangle a}(u-\tau) \left[\ln^2\left(\frac{\tau}{2r_0}\right)+ \frac{1243}{420} \ln\left(\frac{\tau}{2r_0}\right) \right]\nn\\
&+ \frac{192}{175}\,\dM  \int_0^{+\infty} \dd\tau\, \dMbar^{(1)}_{a \langle i} \dMbar^{(6)}_{j \rangle a}(u-\tau) \left[\ln^2\left(\frac{\tau}{2r_0}\right)+ \frac{186743}{15120} \ln\left(\frac{\tau}{2r_0}\right) \right]\nn\\
&+ \frac{1084}{315}\,\dM  \int_0^{+\infty} \dd\tau  \ln\left(\frac{\tau}{2r_0}\right)  \dMbar^{(2)}_{a \langle i}\dMbar^{(5)}_{j \rangle a}(u-\tau)\nn\\
&- \frac{104}{5}\,\dM \int_0^{+\infty} \dd\tau  \ln\left(\frac{\tau}{2r_0}\right) \dMbar^{(3)}_{a \langle i} \dMbar^{(4)}_{j \rangle a}(u-\tau) \nn\\
& + \frac{4}{7} \,\dM \,\dMbar_{a \langle i} \int_0^{+\infty} \dd\tau \, \dMbar^{(7)}_{j \rangle a}(u-\tau) \left[  \ln\left(\frac{\tau}{2r_0}\right) +\frac{15667}{10500} \right]\nn\\
& - \frac{20}{7} \,\dM \,\dMbar^{(1)}_{a \langle i} \int_0^{+\infty} \dd\tau \, \dMbar^{(6)}_{j \rangle a}(u-\tau) \left[  \ln\left(\frac{\tau}{2r_0}\right) -\frac{3590791}{472500} \right]\nn\\
& - \frac{16}{7} \,\dM \,\dMbar^{(2)}_{a \langle i} \int_0^{+\infty} \dd\tau \, \dMbar^{(5)}_{j \rangle a}(u-\tau) \left[  \ln\left(\frac{\tau}{2r_0}\right) + \frac{981149}{94500} \right]\nn\\
& - \frac{3901382}{165375} \dM \, \dMbar^{(3)}_{a \langle i}\dMbar^{(3)}_{j \rangle a}\, .
\end{align}
There is only one doubly-integrated term left, \emph{cf.} the first line of~\eqref{eq:UijHarmMxMijxMijToMv3}, which can be seen as the tails-of-memory contribution properly speaking, whereas the structure of the other singly-integrated terms are more akin to simpler tails-of-tails. 
\newpage
The terms proportional to $\ln^2(\frac{\tau}{2r_0})$ are exactly compensated by those arising in the tail part of $\mathcal{U}_{ij}^{\dM\times\dMbar_{ij} \times \dMbar_{ij}}$ given by~\eqref{subeq:UijRadMxMbarijxMbarijTail}. Notice that not all possible terms allowed by a dimensionality argument are present in~\eqref{eq:UijHarmMxMijxMijToMv3}: for instance there is no term 
\begin{align}
	\propto \dM \,\dMbar_{a\langle i}^{(4)}(u) \int_0^{+\infty} \dd\tau \ln\left(\frac{\tau}{2r_0}\right)\dMbar_{j \rangle a}^{(3)}(u-\tau)\,,
\end{align}
which can be understood from the structure of the cubic source term \eqref{eq:complicatedStructureSource}  corresponding to  the diagram of Fig.~\ref{subfig:Feynman1}.

\subsection{Testing the integration method}
\label{subsubsec:simplifcationMethodTest}

The simplication method in Sec~\ref{subsec:simplifcationMethod} is also very useful to test our integration method. Indeed, let us consider the integration of a typical tail-of-memory source term, but which we multiply by the factor $B$. For simplicity, we choose $k=3$, $m=0$ and $\ell=0$:
\begin{equation}
\label{eq:defNZIntegral}
\mathcal{I}\equiv \FPprop \biggl[ B \left(\frac{r}{r_0}\right)^B \frac{G(t-r)}{r^3}\int_1^{+\infty} \dd x\, Q_0(x) F(t-rx)\biggr]\,.
\end{equation}
This can be computed asymptotically when $r\to+\infty$ using the arsenal of techniques developed in Sec.~\ref{sec:techniques}, and we find
\begin{subequations}
		\label{subeq:expressionNZIntegralv1}
		\begin{align}
		&\mathcal{I} = \frac{1}{r}\biggl[ \Delta - \frac{1}{2} \int_0^{+\infty} \dd \tau \ln\left(\frac{\tau}{2 r_0}\right)G(u-\tau) F^{(1)}(u-\tau)  \biggr] + o\left(\frac{1}{r}\right)\,,
\end{align}
with the rather cumbersome combination of the functionals~\eqref{subeq:defPsibar}:
\begin{align}
		\Delta \equiv -\frac{1}{3}\,\underset{1,2}{\overline{\Psi}_0}[F^{(2)},G] +\frac{1}{3}\,\underset{1,0}{\overline{\Psi}_0}[F^{(2)},G] + \underset{1,1}{\overline{\Psi}_0}[F^{(1)},G^{(1)}] - \underset{1,0}{\overline{\Psi}_0}[F^{(1)},G^{(1)}]\,.
\end{align}
\end{subequations}
But, on the other hand, the presence of the factor $B$ has the effect of only selecting the pole in the $B$-expansion of the integrated source. This allows us to perform first the near-zone expansion $r\rightarrow 0$ of the source, and then to integrate term by term this expansion using the formulae in~\cite{MQ4PN_jauge, TLB22}. Performing the integration in this manner, we find instead 
\begin{equation}
\label{subeq:expressionNZIntegralv2}
\mathcal{I} = -\frac{G(u)}{r} \bigg[F(u) + \int_0^{+\infty} \dd \tau \ln\left(\frac{\tau}{2 r_0}\right) F^{(1)}(u-\tau) \biggr] + o\left(\frac{1}{r}\right)\,.
\end{equation}

The two expressions~\eqref{subeq:expressionNZIntegralv1} and~\eqref{subeq:expressionNZIntegralv2} must be identical, therefore we have found a non-trivial relationship  between the functionals $_{k,m}{\overline{\Psi}_\ell}[F,G]$ which must absolutely be satisfied if our integration method is correct. Applying the simplification method described in Sec~\ref{subsec:simplifcationMethod}, we can prove that these two expressions are indeed identical. This constitutes a strong test confirming simultaneously the soundness of our integration method and of our simplification method. We have repeated this test with many other such sources; all were successful.

\section{Results}
\label{sec:results}

We present our final result, which is the radiative quadrupole moment $\mathcal{U}_{ij}$ parametrizing the asymptotic waveform, expressed in terms of the harmonic canonical moments up \mbox{to 4.5PN order}, including the tail-of-memory contribution. 

Recall that in the previous sections, we have only worked in the radiative construction, and hence with radiative canonical moments denoted $\dMbar_L$ and $\dSbar_L$. This is perfectly legitimate, but until now, all the results in the literature~\cite{BD92,B98tail,B98quad,FMBI12,MBF16} were presented in terms of the harmonic construction and hence of harmonic canonical moments $\dM_L$ and $\dS_L$. Even more crucially, it is the harmonic canonical moments that were computed in previous works in terms of source parameters~\cite{MHLMFB20,MQ4PN_IR,MQ4PN_renorm,MQ4PN_jauge}. For these reasons, we have to express our final result in terms of the harmonic canonical moments $\dM_L$ and $\dS_L$. At 4PN order, this can be done thanks to the moment redefinition derived in~\cite{TLB22}, namely
\begin{equation}\label{eq:momentRedefinition}
	\begin{aligned}
		\dMbar_{ij} =
		\dM_{ij}
		&\
		+ \frac{G\,\dM}{c^3}\,\dM_{ij}^{(1)}  \left[-\frac{26}{15}+2 \ln\left(\frac{r_0}{b_0}\right)\right]+ \frac{G^2\dM^2}{c^6}\,\dM_{ij}^{(2)} \left[\frac{124}{45}-\frac{52}{15} \ln\left(\frac{r_0}{b_0}\right) +2  \ln^2\left(\frac{r_0}{b_0}\right)  \right]\\
		& \
		+ \frac{G^2 \dM}{c^8}
		\bigg[-\frac{8}{21}\,\dM^{}_{a\langle i}\dM_{j\rangle a}^{(4)}
		-\frac{8}{7}\,\dM^{(1)}_{a\langle i}\dM_{j\rangle a}^{(3)}
		-\frac{8}{9}\,\dM^{(3)}_{a\langle i}\dS_{j\rangle \vert a}^{}\bigg]+\mathcal{O}\left(\frac{1}{c^9}\right)\,,
	\end{aligned}
\end{equation}
where $b_0$ is the arbitrary gauge constant introduced in Eq.~\eqref{eq:xi1}, and where $c$ is henceforth reintroduced. The moment redefinition~\eqref{eq:momentRedefinition} should be inserted into the linear and quadratic contributions to the radiative quadrupole written in terms of the radiative metric~\eqref{eq:UijRadQuadratic}, so as to yield a correction at cubic order. 

This correction should be understood in the following sense: let $\mathcal{U}_{ij}^{\text{cubic,rad}} [\dMbar_L,\dSbar_L]$ be the functional expression for the cubic part of the radiative quadrupole in terms of the radiative moments, as worked out in the present paper. Then its counterpart in terms of the harmonic moments, simply denoted $\mathcal{U}_{ij}^{\text{cubic}} [\dM_L,\dS_L]$, is given by
\begin{equation}
	\mathcal{U}_{ij}^{\text{cubic}} [\dM_L,\dS_L]=\mathcal{U}_{ij}^{\text{cubic,rad}} [\dM_L,\dS_L] + \delta\mathcal{U}_{ij}^{\text{cubic,rad}}[\dM_L,\dS_L]\,,
\end{equation}
where $\mathcal{U}_{ij}^{\text{cubic,rad}} [\dM_L,\dS_L]$ means that we simply substituted the radiative moments with the harmonic moments in the radiative functional, and where $\delta\mathcal{U}_{ij}^{\text{cubic,rad}}[\dM_L,\dS_L]$ represents the correction to be applied due to the moment redefinition~\eqref{eq:momentRedefinition}. We find 
 \begin{align}
\label{subeq:UijHarmMxMijxMijCorr}
	\delta\mathcal{U}_{ij}^{\text{cubic,rad}}  &=  \frac{G^2 \dM^2}{c^6} \Bigg\{\left[- \frac{52}{15}+4\ln\left(\frac{r_0}{b_0}\right)\right]\int_0^{+\infty} \dd \tau\, \ln\left(\frac{c \tau}{2 r_0}\right)\dM_{ij}^{(5)}(u-\tau) \nn\\
	&\qquad\qquad + \left[-\frac{257}{75}+\frac{11}{3}\ln\left(\frac{r_0}{b_0}\right)+2\ln^2\left(\frac{r_0}{b_0}\right)\right]\dM_{ij}^{(4)}\Bigg\}\nn\\
	&+ \frac{G^2\dM}{c^8}\Biggl\{\left[-\frac{16}{105}- \frac{8}{7}\ln\left(\frac{r_0}{b_0}\right)\right] \dM_{a \langle i}^{(3)}  \dM_{j \rangle a}^{(3)} +\left[- \frac{14}{15}- 2 \ln\left(\frac{r_0}{b_0}\right)\right]\dM_{a \langle i}^{(2)}  \dM_{j \rangle a}^{(4)}\nn \\
	& \qquad\qquad  
	+\left[-  \frac{32}{35} - \frac{8}{7}\ln\left(\frac{r_0}{b_0}\right) \right]\dM_{a \langle i}^{(1)}  \dM_{j \rangle a}^{(5)} + \left[ - \frac{22}{35} + \frac{2}{7}\ln\left(\frac{r_0}{b_0}\right) \right]\dM_{a \langle i}  \dM_{j \rangle a}^{(6)}\Biggr\}\nn\\
	&+ \frac{G^2\dM}{c^8}\dS_{a\vert \langle i}\left[\frac{22}{15} - \frac{2}{3}\ln\left(\frac{r_0}{b_0}\right)\right] \dM_{j \rangle a}^{(5)}\,.
\end{align}

We finally present the full radiative quadrupole, parametrized by the harmonic canonical moments $\dM_L$ and $\dS_L$, including all quadratic, cubic and quartic contributions that contribute up to the 4.5PN order. It reads
\begin{equation}
\mathcal{U}_{ij} = \mathcal{U}_{ij}^{\text{linear}} + \,\mathcal{U}_{ij}^{\text{quadratic}} + \,\mathcal{U}_{ij}^{\text{cubic}} + \,\mathcal{U}_{ij}^{\text{quartic}}\,,
\end{equation}
where the linear contribution is just $\mathcal{U}_{ij}^{\text{linear}}=\dM_{ij}^{(2)}$, the quadratic part $\mathcal{U}_{ij}^{\text{quadratic}}$ can be entirely found in Eqs.~(4.4-5-6) of~\cite{FMBI12}, the quartic piece $\mathcal{U}_{ij}^{\text{quartic}}$ is given in Eq.~(4.8) of~\cite{MBF16}, and where the cubic contributions
\begin{equation}
	\mathcal{U}_{ij}^{\text{cubic}} =  \mathcal{U}_{ij}^{\dM^2\times\dM_{ij}}+\mathcal{U}_{ij}^{\dM\times\dS_i\times\dM_{ij}}+\mathcal{U}_{ij}^{\dM\times\dM_{ij}\times\dM_{ij}}\,,
\end{equation}
decompose into the tail-of-tail~\cite{B98tail,FBI15}, and, both new with this paper, the spin-quadrupole tail and the tail-of-memory. We have
\begin{subequations}\label{eq:Uijcubic}
\begin{align}
\label{subeq:UijHarmMxMxMij}
	\mathcal{U}_{ij}^{\dM^2 \times \dM_{ij}} &= \frac{2 G^2\dM^2}{c^6}\!\!\int_{0}^{+\infty} \!\!\!\dd\tau\,\dM_{ij}^{(5)}(u-\tau) \! \left[\ln^2\left(\frac{c \tau}{2b_0}\right) + \frac{11}{6} \ln\left(\frac{c \tau}{2b_0}\right)- \frac{107}{105} \ln\left(\frac{c \tau}{2r_0}\right) + \frac{124627}{44100}\right]\! ,\\
	\label{subeq:UijHarmMxSixMij}
	\mathcal{U}_{ij}^{\dM\times\dS_i \times \dM_{ij}} &= - \frac{2 G^2 \dM}{ 3 c^8} \,\dS_{a\vert \langle i} \int_{0}^{+\infty} \dd\tau\,  \dM_{j\rangle a}^{(6)}(u-\tau) \left[ \ln\left(\frac{c \tau}{2b_0}\right) +2 \ln\left(\frac{c \tau}{2r_0}\right)  + \frac{1223}{1890}  \right]\, ,\\
	\label{subeq:UijHarmMxMijxMij}
\mathcal{U}_{ij}^{\dM\times\dM_{ij} \times \dM_{ij}} &= \frac{8 G^2\dM}{7 c^8}\Bigg\{\int_0^{+\infty} \!\dd\rho\,  \dM_{a \langle i}^{(4)}(u-\rho) \int_0^{+\infty} \!\dd \tau\,  \dM_{j \rangle a}^{(4)}(u-\rho-\tau)  \left[ \ln\left(\frac{c \tau}{2 r_0}\right) - \frac{1613}{270} \right]\nn\\
&\quad - \frac{5}{2} \int_0^{+\infty} \!\dd\tau \,  (\dM^{(3)}_{a \langle i} \dM^{(4)}_{j \rangle a})(u-\tau)\left[\ln\left(\frac{c \tau}{2 r_0}\right)+\frac{3}{2} \ln\left(\frac{c \tau}{2b_0}\right) \right]\nn\\
&\quad - 3  \int_0^{+\infty} \!\dd\tau\, (\dM^{(2)}_{a \langle i}  \dM^{(5)}_{j \rangle a})(u-\tau) \left[\ln\left(\frac{c \tau}{2r_0}\right)  +\frac{11}{12} \ln\left(\frac{c \tau}{2b_0}\right)  \right]\nn\\
&\quad-\frac{5}{2} \int_0^{+\infty} \!\dd\tau \, (\dM^{(1)}_{a \langle i}\dM^{(6)}_{j \rangle a})(u-\tau) \left[\ln\left(\frac{c \tau}{2r_0}\right)  + \frac{3}{10} \ln\left(\frac{c \tau}{2 b_0}\right) \right]\nn\\
&\quad - \int_0^{+\infty} \!\dd\tau\,(\dM_{a \langle i}\dM^{(7)}_{j \rangle a})(u-\tau) \left[\ln\left(\frac{c \tau}{2r_0}\right) - \frac{1}{4}\ln\left(\frac{c \tau}{2 b_0}\right)\right]\nn\\
&\quad - 2  \dM^{(2)}_{a \langle i} \int_0^{+\infty} \!\dd\tau\,  \dM^{(5)}_{j \rangle a}(u-\tau) \left[ \ln\left(\frac{c \tau}{2r_0}\right)+ \frac{27521}{5040} \right]\nn\\
&\quad- \frac{5}{2}\,  \dM^{(1)}_{a \langle i} \int_0^{+\infty} \!\dd\tau \, \dM^{(6)}_{j \rangle a}(u-\tau)  \left[\ln\left(\frac{c \tau}{2r_0}\right)+\frac{15511}{3150} \right]\nn\\
&\quad+ \frac{1}{2} \, \dM_{a \langle i} \int_0^{+\infty} \!\dd\tau \,\dM^{(7)}_{j \rangle a}(u-\tau)  \left[ \ln\left(\frac{c \tau}{2r_0}\right)  - \frac{6113}{756}\right]\,  \Bigg\}\,. 
\end{align}
\end{subequations}
The double integral in  the first line of \eqref{subeq:UijHarmMxMxMij} is the genuine tail-of-memory. However, when computing the flux, we must actually compute the time-derivative of the radiative quadrupole. We then find that the tail-of-memory term becomes a simple tail term in the flux, in the same manner as the quadrupole-quadrupole memory term becomes instantaneous in the flux \cite{B98quad}. 
\newpage
We can perform three important tests of our end result:
\begin{itemize}
\item The constant $b_0$ can be eliminated at the level of the full radiative quadrupole at 4PN order by the shift in the time coordinate
\begin{align}
	u' = u -\frac{2 G \dM}{c^3}\ln b_0 \,,
\end{align}
along with a Taylor expansion of the canonical quadrupole moment. This shows that $b_0$ is just associated to a choice for the origin of time in the asymptotic radiative coordinate system and is clearly unphysical.
\item When isolating the contribution of $\ln r_0$ to the radiative quadrupole for the three cubic interactions~\eqref{eq:Uijcubic}, we find
\begin{align}
	\mathcal{U}_{ij}^\text{cubic} &=  \ln r_0 \bigg[\frac{214}{105}\frac{G^2 \dM^2}{c^6} \dM_{ij}^{(4)} \nn\\
	& \quad\quad + \frac{G^2 M}{c^8}\Big( 4  \dM_{a \langle i}^{(2)}\dM_{j \rangle a}^{(4)}  + \frac{32}{7}  \dM_{a \langle i}^{(1)}\dM_{j \rangle a}^{(5)} + \frac{4}{7}  \dM_{a \langle i} \dM_{j \rangle a}^{(6)} + \frac{4}{3} \dS_{a \vert \langle  i} \dM_{j \rangle a}^{(5)} \Big) \bigg]\nn \\
	&\quad\quad + (\text{terms independent of $r_0$}) \,.
\end{align}
This expression exactly cancels the $\ln r_0$ terms arising in the second time derivative of Eq.~(6.1) in~\cite{MQ4PN_renorm}, which accounts for the contribution of the dimensional regularization of the cubic interactions to the renormalized canonical quadrupole moment for compact binaries. This cancellation of the $\ln r_0$ terms with those in the end result of~\cite{MQ4PN_renorm} is a strong indication of the correctness of our results~\eqref{eq:Uijcubic}.

\item In the MPM approach, the leading-order quadratic memory term, see \emph{e.g.} (5.10) in \cite{B98quad}, reads
\begin{equation}\label{eq:standardMemory}
	\mathcal{U}_{ij}^{\text{mem}}= - \frac{2G}{7c^5}\int_0^{+\infty} \dd \rho \, \dM_{a \langle i}^{(3)}(u-\rho)\dM_{j \rangle a}^{(3)}(u-\rho) + \mathcal{O}(G^2)\,.
\end{equation}
We find here that the genuine tail-of-memory given by the first term of \eqref{subeq:UijHarmMxMijxMij} can be obtained simply by replacing in \eqref{eq:standardMemory} the canonical quadrupole moment $\dM_{ij}$ by the radiative quadrupole moment itself, including the dominant tail effect, \emph{i.e.}
\begin{equation}\dM_{ij}^{(2)} \longrightarrow \mathcal{U}_{ij} = \dM_{ij}^{(2)} + \frac{2G \dM}{c^3} \int_0^{+\infty} \dd\tau \dM_{ij}^{(4)} \left[\ln\left(\frac{c \tau}{2b_0}\right)+\frac{11}{12}\right] + \mathcal{O}\left(\frac{1}{c^5}\right)\,,\end{equation}
along with a reexpansion at cubic order and an integration by parts (the constant $11/12$ is irrelevant here). With our result~\eqref{subeq:UijHarmMxMijxMij}, we thus explicitly retrieve at this order the expression of memory effects computed using the radiative moments defined at future null infinity, see \emph{e.g.}~\cite{F09,F11}.

\end{itemize}
\newpage
\acknowledgments

The authors thank Fran\c{c}ois Larrouturou for useful suggestions and interesting discussions at different stages of this project. We also thank Geoffrey Compère for pointing out some typos in the quadratic metric, as well as Laura Bernard, Guillaume Faye, Quentin Henry and Stavros Mougiakakos for interesting discussions. L.B. acknowledges the Institut de Physique Théorique (IPhT) in CEA/Saclay for a visiting position. D.T.~thanks Laura Bernard and the Kavli Institute for Theoretical Physics (KITP) in Santa Barbara for their invitation to participate to the ``High-Precision Gravitational Waves'' program (supported in part by the National Science Foundation under Grant No. NSF PHY-1748958).


\appendix


\section{Extracting $N_L$, $P_L$, $Q_L$ and $R_L$}
\label{app:NPQR}

We describe the practical method we use to extract four sets of STF moments $N_L$, $P_L$, $Q_L$ and $R_L$ parametrizing a vector quantity satisfying $\Box w^\mu = 0$. Here $w^\mu\equiv w_n^\mu$ represents the divergence $w_n^\mu=\partial_\nu u_n^{\mu\nu}$ of the quantity~\eqref{eq:unMPM} following the MPM algorithm. Having extracted these moments we can then construct the tensor $v_n^{\mu\nu} \equiv  \mathcal{V}^{\mu\nu}[N_L, P_L, Q_L, R_L]$ using the formulae~(48) in~\cite{BlanchetLR}, which satisfies at once $\Box v_n^{\mu\nu}~=~0 $ and $\partial_\nu v_n^{\mu\nu}=- w_n^\mu$. By definition, see~(47) in~\cite{BlanchetLR},

\begin{subequations}\label{eq:WparamNPQR}
\begin{align}
		w^0 &= \sum_{\ell=0}^{+\infty} \partial_L\left[ r^{-1} N_L(t-r)\right]\,,\\
		w^i &= \sum_{\ell=0}^{+\infty} \partial_{iL}\left[ r^{-1} P_L(t-r)\right] \nn\\& + \sum_{\ell=1}^{+\infty} \partial_{L-1}\left[ r^{-1} Q_{iL}(t-r)\right] +  \sum_{\ell=1}^{+\infty} \epsilon_{iab} \partial_{aL-1}\left[ r^{-1} R_{bL-1}(t-r)\right]\,.
\end{align}
\end{subequations}
Next, we define the auxiliary quantities
\begin{subequations}
\begin{align}
\widetilde{w}^i  &\equiv  w^i -  \sum_{\ell=0}^{+\infty} \partial_{iL}\left[ r^{-1} P_L(t-r)\right] \nn\\ &= \sum_{\ell=1}^{+\infty} \partial_{L-1}\left[ r^{-1} Q_{iL}(t-r)\right] +  \sum_{\ell=1}^{+\infty} \epsilon_{iab} \partial_{aL-1}\left[ r^{-1} R_{bL-1}(t-r)\right]\,,\\
\overset{\approx}{w}{}^i &\equiv \widetilde{w}^i    -  \sum_{\ell=1}^{+\infty} \partial_{L-1}\left[ r^{-1} Q_{iL}(t-r)\right] =  \sum_{\ell=1}^{+\infty} \epsilon_{iab} \partial_{aL-1}\left[ r^{-1} R_{bL-1}(t-r)\right]\,.
\end{align}
\end{subequations}
Using formulae from Appendix~A in~\cite{BD86}, we express the angular integrals 
\begin{subequations}
\begin{align}
\int \frac{\dd\Omega}{4\pi}\,\hat{n}_L w^0&= \frac{\ell!}{(2\ell+1)!!}\,r^\ell \left(\frac{1}{r}\frac{\partial}{\partial r}\right)^\ell \bigg[r^{-1}N_L(t-r)\bigg]\,,\\
\int \frac{\dd\Omega}{4\pi}\,\hat{n}_{iL} w^i&= \frac{(\ell+1) \ell!}{(2\ell+1)(2\ell+1)!!}\,r^{\ell+1} \left(\frac{1}{r}\frac{\partial}{\partial r}\right)^{\ell+1} \bigg[r^{-1}P_L(t-r)\bigg]\,,\\
\int \frac{\dd\Omega}{4\pi}\,n_i\hat{n}_{L} \widetilde{w}^i&= \frac{\ell!}{(2\ell+1)!!}\,r^{\ell-1} \left(\frac{1}{r}\frac{\partial}{\partial r}\right)^{\ell-1} \bigg[r^{-1}Q_L(t-r)\bigg]\,,\\
\epsilon_{ab\langle i}\!\int \frac{\dd\Omega}{4\pi}\,n_{L-1\rangle}n_b \overset{\approx}{w}{}^a&= \frac{(\ell+1)(\ell-1)!}{(2\ell+1)!!}\,r^{\ell} \left(\frac{1}{r}\frac{\partial}{\partial r}\right)^{\ell} \bigg[r^{-1}R_{iL-1}(t-r)\bigg]\,.
\end{align}
\end{subequations}
If $w^\mu$ is known exactly to all order in $r$, we find that the multipole moments can be computed using the near zone limit $r\to 0$ (with $t-r=$ const) as
\begin{subequations}
\begin{align}
N_L(t-r) &= \frac{(-)^\ell(2\ell+1)}{\ell!} \lim_{r\rightarrow 0}\bigg[r^{\ell+1}\int \frac{\dd\Omega}{4\pi}\, \hat{n}_L w^0\bigg]\,,\\
P_L(t-r) &= \frac{(-)^{\ell+1}(2\ell+1)}{(\ell+1)!} \lim_{r\rightarrow0}\bigg[r^{\ell+2}\int \frac{\dd\Omega}{4\pi}\, \hat{n}_{iL} w^i\bigg]\,,\\
Q_L(t-r) &= \frac{(-)^{\ell-1}(2\ell+1)(2\ell-1)}{\ell!} \lim_{r\rightarrow0}\bigg[r^{\ell}\int \frac{\dd\Omega}{4\pi}\, n_i \hat{n}_{L} \widetilde{w}^i\bigg]\,,\\
R_{iL-1}(t-r) &= \frac{(-)^{\ell}(2\ell+1)}{(\ell+1)(\ell-1)!} \lim_{r\rightarrow0}\bigg[r^{\ell+1}\int \frac{\dd\Omega}{4\pi}\, \epsilon_{ab\langle i} n_{L-1\rangle} n_b \overset{\approx}{w}{}^{a}\bigg]\,.
\end{align}
\end{subequations}
If we only know the leading order of the asymptotic expansion of $w^\mu$ as $r\to+\infty$, the previous expressions cannot be used. In that case, we get equivalent expressions for the time derivatives of the multipole moments:
\begin{subequations}
\begin{align}
\overset{(\ell)}{N_L}(t-r) &= \frac{(-)^\ell(2\ell+1)!!}{\ell!} \lim_{r\to+\infty}\bigg[r \int \frac{\dd\Omega}{4\pi}\, \hat{n}_L w^0\bigg]\,,\\
\overset{(\ell+1)}{P_L}(t-r) &= \frac{(-)^{\ell}(2\ell+1)(2\ell+1)!!}{(\ell+1)!} \lim_{r\to+\infty}\bigg[r \int \frac{\dd\Omega}{4\pi}\, \hat{n}_{iL} w^i\bigg]\,,\\
\overset{(\ell-1)}{Q_L}(t-r) &= \frac{(-)^{\ell-1}(2\ell+1)!!}{\ell!} \lim_{r\to+\infty}\bigg[r \int \frac{\dd\Omega}{4\pi}\, n_i \hat{n}_{L} \widetilde{w}^i\bigg]\,,\\
\overset{(\ell)}{R}_{iL-1}(t-r) &= \frac{(-)^{\ell}(2\ell+1)!!}{(\ell+1)(\ell-1)!} \lim_{r\to+\infty}\bigg[r \int \frac{\dd\Omega}{4\pi}\, \epsilon_{ab\langle i} n_{L-1\rangle} n_b \overset{\approx}{w}{}^a\bigg]\,.
\end{align}
\end{subequations}
\section{Proof of convergence when $k=1$ and $k=2$}
\label{app:convergence}

In this Appendix, we assume that $k\in\{1,2\}$. We shall prove that in these cases one can set $B=0$ in the general solution~\eqref{eq:expressionPsiBkml}, for arbitrary smooth functions $F$ and $G \in\mathcal{C}^\infty(\mathbb{R})$ which vanish identically in the remote past, for $t\leqslant-\mathcal{T}$. Although it should be possible to check this point using the expression of the three-dimensional retarded integral of the source term~\eqref{eq:sourceBToM}, we present here a detailed analysis directly based on the structure of the solution~\eqref{eq:expressionPsiBkml}.

 We must first do some manipulation so as to transfer the boundary conditions of  $F$ and $G$ into the bounds of the integrals. Thanks to the regularization factor $\lambda^B$, we can always manipulate these integrals safely, since they are defined by analytic continuation for any $B\in\mathbb{C}$ except at some integer values including the value of interest $B=0$, at which we apply the finite part in the end.
\subsubsection{Case $k+j \leqslant 1$}
In this case, we can perform the change of variable $\lambda = \mu + \rho/2$ on Eq.~\eqref{subeq:expressionPhiBkmij} and expand using the binomial formula:
\begin{align} 
	\phi_{ij}^B(u,r) &= \sum_{p=0}^{-k-j+1} \begin{pmatrix}-k-j+1\\p\end{pmatrix} \int_{0}^{+\infty}\dd \rho\, \rho^{i+j}\left( \frac{\rho}{2}\right)^{-k-j+1-p} G(u-\rho)  \int_1^{+\infty} \dd x \, Q_m(x)\nn \\
&  \qquad\qquad\qquad \times \int_{0}^{r}\dd\mu\, \left(\frac{\mu+\rho/2}{r_0}\right)^B \mu^p \,F\left(u-\frac{\rho(x+1)}{2}-\mu(x-1)\right)\,.
\end{align}
We then set $\tau = \mu(x-1)$ with $\dd \tau = (x-1)\dd \mu$, invert the integrals, and find
\begin{align} 
	\phi_{ij}^B(u,r) &= \sum_{p=0}^{-k-j+1} 2^{-k+j-1+p}\begin{pmatrix}-k-j+1\\p\end{pmatrix}  \int_{0}^{+\infty}\dd\tau\,  \tau^p   \int_{1+\tau/r}^{+\infty} \dd x \, \frac{Q_m(x)}{(x-1)^{p+1}} 
\left(\frac{\frac{\tau}{x-1}+\rho/2}{r_0}\right)^B \nn\\&  \qquad\qquad\qquad\times\int_{0}^{+\infty}\dd \rho\, \rho^{-k+i+1-p}G(u-\rho) F\left(u-\frac{\rho(x+1)}{2}-\tau \right)\,. 
\end{align}
Next we introduce 
\begin{equation}
	\Upsilon_{k,i,j,p}(\tau) \equiv  \tau^p   \int_{1+\tau/r}^{+\infty} \dd x \, \frac{Q_m(x)}{(x-1)^{p+1}}\int_{0}^{+\infty}\dd \rho\, \rho^{-k+i+1-p}G(u-\rho) F\left(u-\frac{\rho(x+1)}{2} - \tau \right)\,,
\end{equation}
which is well defined, since $-k+i+1-p\geqslant i+j\geqslant 0$ ensures that the integral in $\rho$ converges at $0$, and the behavior of $Q_m(x)/(x-1)^{p+1} \sim x^{-m-p-2}$ when $x\to+\infty$ ensures that the integral in $x$ converges at infinity.

Since $F$ vanishes identically in the remote past, so does $\Upsilon_{k,i,j,p}(\tau)$, thus it is clearly integrable at $\tau \rightarrow \infty$. We then bound this quantity with
\begin{equation}
	\left| \Upsilon_{k,i,j,p}(\tau) \right| \leqslant  \tau^p\int_{1+\tau/r}^{+\infty}  \dd x \,  \frac{Q_m(x)}{(x-1)^{p+1}} = \mathcal{O}_{\tau\rightarrow 0}\left(\ln^2 \tau\right)\,,
\end{equation}
which proves integrability at the bound $\tau \rightarrow 0$.

\subsubsection{Case $k+j \geqslant  2$}

We first shuffle the order of the integrals in Eq.~\eqref{subeq:expressionPhiBkmij} and obtain
\begin{align}\label{PhiReshuffled}
\phi_{ij}^B &=    \int_1^{+\infty} \dd x \,Q_m(x)  \int_{0}^{+\infty}\dd \lambda\, \left(\frac{\lambda}{r_0}\right)^B \lambda^{-k-j+1}  \int_0^{2\lambda} \dd\rho\,    \, \rho^{i+j}G(u-\rho)  F\bigl[u-\rho-\lambda(x-1)\bigr] \nn\\
& -    \int_1^{+\infty} \dd x \, Q_m(x)  \int_{r}^{+\infty}\dd \lambda\, \left(\frac{\lambda}{r_0}\right)^B   \lambda^{-k-j+1}\int_0^{2(\lambda-r)} \dd\rho\,    \, \rho^{i+j}G(u-\rho)  F\bigl[u-\rho-\lambda(x-1)\bigr]\,.
\end{align}
\newpage
Taking advantage of the conditions on $F$ and $G$, we can now restrict the bounds of the integrals to $\lambda \leqslant \frac{u+\mathcal{T}}{x-1}$. This yields
\begin{align}\label{PhiReshuffledBounded} 
	\phi_{ij}^B &=    \int_1^{+\infty} \dd x \,Q_m(x)  \int_{0}^{\frac{u+\mathcal{T}}{x-1}}\dd \lambda\, \left(\frac{\lambda}{r_0}\right)^B \lambda^{-k-j+1}  \int_0^{\min(u+\mathcal{T},\,2\lambda)} \dd\rho\,    \, \rho^{i+j}G(u-\rho)  F\bigl[u-\rho-\lambda(x-1)\bigr] \nn\\
& -    \int_1^{\frac{u+\mathcal{T}}{r}} \dd x \, Q_m(x)  \int_{r}^{\frac{u+\mathcal{T}}{x-1}}\dd \lambda\, \left(\frac{\lambda}{r_0}\right)^B   \lambda^{-k-j+1}\int_0^{\min(u+\mathcal{T},\,2(\lambda-r))} \dd\rho\,    \, \rho^{i+j}G(u-\rho)  F\bigl[u-\rho-\lambda(x-1)\bigr]\,.
\end{align}
We introduce 
\begin{equation}
\Xi_{k,i,j}(x) \equiv \int_{0}^{\frac{u+\mathcal{T}}{x-1}}\dd \lambda\, \lambda^{-k-j+1}  \int_0^{\min(u+\mathcal{T},\,2\lambda)} \dd\rho\,    \, \rho^{i+j}G(u-\rho)  F\bigl[u-\rho-\lambda(x-1)\bigr]\,.
\end{equation}
Let us bound this quantity for any $x\in\  ]1,\infty]$:
\begin{align}
\big|\Xi_{k,i,j}(x)\big| &\leqslant  \int_{0}^{\frac{u+\mathcal{T}}{x-1}}\dd \lambda\, \lambda^{-k-j+1}  \int_0^{\min(u+\mathcal{T},\,2\lambda)} \dd\rho\,    \, \rho^{i+j}\Big|G(u-\rho)  F(u-\rho-\lambda(x-1))  \Big|\\
&\leqslant \mathcal{K} \int_{0}^{\frac{u+\mathcal{T}}{x-1}}\dd \lambda\, \lambda^{-k-j+1}  \int_0^{\min(u+\mathcal{T},\,2\lambda)} \dd\rho\,    \, \rho^{i+j}\nn\\
&= \frac{\mathcal{K}}{i+j+1}  \int_{0}^{\frac{u+\mathcal{T}}{x-1}}\dd \lambda\, \lambda^{-k-j+1}\Big[ \min(u+\mathcal{T},\,2\lambda) \Big]^{i+j+1}\nn\\
& = \frac{\mathcal{K}}{i+j+1}\Bigg\{\int_{0}^{\frac{u+\mathcal{T}}{2}}\dd \lambda\, \lambda^{-k-j+1}(2\lambda) ^{i+j+1}+\theta(3-x)\int_{\frac{u+\mathcal{T}}{2}}^{\frac{u+\mathcal{T}}{x-1}}\dd \lambda\, \lambda^{-k-j+1}(u+\mathcal{T}) ^{i+j+1}\Bigg\}\nn\\
& = \frac{\mathcal{K}}{i+j+1}\Bigg\{\frac{2^{j+k-2}}{i+(3-k)}(u+\mathcal{T})^{i+(3-k)}+\theta(3-x) (u+\mathcal{T}) ^{i+j+1} \int_{\frac{u+\mathcal{T}}{2}}^{\frac{u+\mathcal{T}}{x-1}}\dd \lambda\, \lambda^{-k-j+1}\Bigg\}\,,\nn
\end{align}
where
\begin{equation}
\mathcal{K}\equiv\sup_{(\tau,\tau')\in[-\mathcal{T},u]^2}\Big[F(\tau)G(\tau')\Big]\,.
\end{equation}
We now distinguish the case $j=2-k$, which yields
\begin{equation}
\big|\Xi_{2,i,0}(x)\big| \leqslant \frac{\mathcal{K}}{i+1}\Bigg\{\frac{1}{i+1}(u+\mathcal{T})^{i+1}+\theta(3-x) (u+\mathcal{T}) ^{i+1}\ln\left(\frac{2}{x-1}\right)\Bigg\}\,,
\end{equation}
and the more general case $k+j\geqslant 2$, in which we find
\begin{equation}
\big|\Xi_{k,i,j}(x)\big| \leqslant \frac{\mathcal{K}(u+\mathcal{T})^{i+(3-k)}}{i+j+1}\Bigg\{\frac{2^{j+k-2}}{i+(3-k)}+\frac{\theta(3-x)}{j+(k-2)}\Big(2^{j+(k-2)}-(x-1)^{k+j-2}\Big)\Bigg\}\,.
\end{equation}

Since $Q_m(x) = \mathcal{O}_{x\rightarrow1^+}\left(\ln\left(x-1\right)\right)$ and $Q_m(x) = \mathcal{O}_{x\to+\infty}\left(x^{-m-1}\right)$, we can now look at the asymptotic behavior of our integrand. In the case  $(k,j)=(2,0)$, we find
\begin{equation}
	Q_m(x)\, \Xi_{2,i,0}(x) = \begin{cases}\displaystyle\calO_{x\rightarrow 1^+}\left(\ln(x-1)^2\right)\,, \\[0.3cm] \displaystyle\calO_{x\to+\infty}\left(x^{-m-1}\right)\,.\end{cases}
\end{equation}
In the general case $k+j\geqslant 2$, we instead have
\begin{equation}
Q_m(x)\, \Xi_{k,i,j}(x)	= \begin{cases}\displaystyle\calO_{x\rightarrow 1^+}\left(\ln(x-1)(x-1)^{j+(k-2)}\right)\,, \\[0.3cm] \displaystyle\calO_{x\to+\infty}\left(x^{-m-1}\right)\,.\end{cases}
\end{equation}

It is now clear that $Q_m(x)\, \Xi_{k,i,j}(x)$ is integrable at both bounds $x\rightarrow 1^+$ and $x\to+\infty$ as soon as $k\in\{1,2\}$. After doing a similar (and much easier) analysis on the second member of Eq.~\eqref{PhiReshuffledBounded}, we find that Eqs.~\eqref{PhiReshuffledBounded}--\eqref{PhiReshuffled}, and consequently Eq.~\eqref{subeq:expressionPhiBkmij}, have a convergent limit as $B\rightarrow 0$, \emph{i.e.}, they do not develop poles in $1/B$. Repeating the analysis of this section with an extra factor $\ln(r/r_0)$ in the integrand shows that we can also safely compute the $1/B$ coefficient in the Laurent series.

\begin{table}
	\caption{Coefficients $\mathcal{A}_{m,\ell}^n$, $\mathcal{B}_{m,\ell}^n$, $\mathcal{C}_{m,\ell}^n$, $\mathcal{D}_{m,\ell}^n$ entering the radiative quadrupole}
	\begin{subtable}[h] {\textwidth} \def\arraystretch{1.25}
		\subcaption{Coefficients $\mathcal{A}_{m,\ell}^{n}$ associated to ${}_{1,m}\overline{\Psi}_\ell [\dMbar_{a\langle i}^{(n)},\dMbar_{j\rangle a}^{(8-n)}] $} 
		\begin{tabular}{|c|c|c|c|c|c|c|c|c|c|c|c|c|c|c|}
			\hline
			$\ell$             & \multicolumn{5}{c|}{2} & \multicolumn{5}{c|}{3} & \multicolumn{4}{c|}{4} \\ \hline
			\diagbox{$m$}{$n$} & 0  & 1  & 2  & 3  & 4  & 0  & 1  & 2  & 3  & 4  & 0  & 1  & 2  & 3   \\ \hline
			0                  & $- \frac{136}{735}$   &  $-\frac{13808}{2625}$  &   $-\frac{34576}{2625}$   &    $\frac{120}{7}$ &   $\frac{16}{7}$  &  $0$   & $\frac{832}{535}$    &   $\frac{29432}{1575}$  &   $\frac{88}{9}$  &   -$\frac{32}{7}$   &  $- \frac{928}{25725}$  &   $-\frac{1168}{3675}$  &  $\frac{664}{525}$   &  $\frac{64}{21}$      \\ \hline
			1                  &  $0$  &   $-\frac{3344}{6125}$  &   $\frac{6184}{875}$  &   $-\frac{16456}{875}$  &  $-\frac{32}{7}$   &  $\frac{1216}{1225}$    &  $\frac{3464}{525}$  &  $-\frac{11584}{525}$   &   $-\frac{17032}{525}$  &   $\frac{64}{7}$ &  $0$  &   $-\frac{9256}{8575}$ &  $-\frac{6256}{1225}$   &   $\frac{3776}{1225}$   \\ \hline
			2                  &   $ \frac{8}{75}$ &  $\frac{18328}{3675}$   &   $\frac{22448}{3675}$  &  $\frac{2528}{175}$   &  $\frac{16}{7}$  &  $0$    &   $-\frac{3776}{735}$  & $\frac{36272}{2205}$   &  $\frac{304}{45}$   &   $-\frac{32}{7}$  &  $- \frac{16}{343}$  &   $-\frac{2896}{5145}$      &  $\frac{1024}{735}$   &  $\frac{4696}{735}$    \\ \hline
			3                  & $0$   & $-\frac{568}{875}$   & $-\frac{1224}{875}$   & $\frac{816}{875}$   & $0$   & $- \frac{128}{175}$    &   $\frac{536}{525}$   & $-\frac{5176}{525}$   & $-\frac{288}{25}$   & $0$   &  $0$  &  $\frac{7064}{3675}$  & $\frac{4616}{1225}$   & $\frac{1464}{1225}$  \\ \hline
			4                  &  $ \frac{96}{275}$  &  $\frac{13152}{6125}$  & $\frac{8544}{6125}$   & $0$   &  $0$  &    $0$  & $-\frac{4608}{1225}$  & $-\frac{3936}{1125}$   & $0$   & $0$   &  $\frac{10336}{94325}$   &  $\frac{4192}{8575}$  &  $-\frac{1616}{1225}$  & $0$    \\ \hline
			5                  &  $0$  &  $-\frac{832}{1225}$  & $0$   & $0$   & $0$   &    $- \frac{64}{245}$  &  $-\frac{32}{105}$    &  $0$  & $0$   &$0$ & $0$   &  $-\frac{464}{1029}$  &  $0$    & $0$ \\ \hline
			6                  &    $- \frac{3648}{13475}$   & $0$    & $0$   & $0$   & $0$   &  $0$   & $0$     & $0$   & $0$   &  $0$   &  $- \frac{304}{11319}$   &  $0$  &  $0$  &  $0$  \\ \hline
		\end{tabular}
	\end{subtable}
	\begin{subtable}{\textwidth}\def\arraystretch{1.25}
		\subcaption{Coefficients $ \mathcal{B}_{m,\ell}^{n}$ associated to $ \,_{2,m}\overline{\Psi}_\ell [\dMbar_{a \langle i}^{(n)},\dMbar_{j \rangle a}^{(7-n)}] $}
		\begin{tabular}{|c|c|c|c|c|c|c|c|c|c|c|c|c|}
			\hline
			$\ell$             & \multicolumn{4}{c|}{2} & \multicolumn{4}{c|}{3} & \multicolumn{4}{c|}{4} \\ \hline
			\diagbox{$m$}{$n$} & 0  & 1  & 2  & 3  & 0  & 1  & 2  & 3   & 0  & 1  & 2  & 3   \\ \hline
			0                  & $0$  &  $\frac{696}{875}$   &  $\frac{312}{7}$  &   $\frac{48}{7}$  & $\frac{608}{175}$   & $\frac{10768}{525}$ &  $- \frac{304}{21}$    & $-\frac{192}{7}$  & $0$   &  $\frac{272}{147}$   &  $\frac{160}{21}$  &   $0$   \\ \hline
			1                  &  $-\frac{15336}{6125}$  & $-\frac{5632}{875}$   & $-\frac{18728}{875}$   & $\frac{64}{7}$   & $0$& $-\frac{5952}{175}$  &  $-\frac{5744}{175}$   & $\frac{176}{7}$  & $-\frac{2432}{1715}$   & $-\frac{1616}{245}$   & $\frac{1040}{49}$  & $\frac{48}{7}$         \\ \hline
			2                  &$0$ &$-\frac{2928}{1225}$  & $\frac{1584}{175}$   & $\frac{48}{35}$    & $-\frac{32}{49}$   &  $\frac{15376}{735}$    &  $-\frac{3424}{105}$   &  $-\frac{240}{7}$  & $0$   & $\frac{3952}{1029}$   &  $\frac{2672}{147}$   & $\frac{48}{7}$   \\ \hline
			3                  &   $-\frac{216}{125}$ &  $\frac{552}{875}$   & $-\frac{2592}{875}$   &  $0$  & $0$   & $-\frac{1728}{175}$  & $-\frac{288}{25}$   & $0$   & $\frac{3904}{2205}$   & $\frac{656}{245}$   & $-\frac{288}{49}$   &  $0$    \\ \hline
			4                  & $0$ &  $-\frac{41472}{6125}$ & $0$ & $0$ &$-\frac{3456}{1225}$  & $-\frac{2592}{1225}$   & $0$   & $0$   & $0$   & $-\frac{608}{343}$   & $0$ & $0$          \\ \hline
			5                  &  $-\frac{1536}{1225}$  & $0$   &  $0$  & $0$   & $0$   & $0$ & $0$    & $0$   & $-\frac{1088}{3087}$   & $0$ &$0$    & $0$   \\ \hline
		\end{tabular}
	\end{subtable}
	
	\begin{subtable}[h]{0.49 \textwidth}\def\arraystretch{1.25}
		\subcaption{Coefficients $ \mathcal{C}_{m,\ell}^{n}$ associated to $ \,_{1,m}\overline{\chi}_\ell [\dMbar_{a \langle i}^{(n)},\dMbar_{j \rangle a}^{(8-n)}] $}
		\begin{tabular}{|c|c|c|c|c|}
			\hline
			$\ell$             & \multicolumn{4}{c|}{2}  \\ \hline
			\diagbox{$m$}{$n$} & 0  & 1  & 2  & 3    \\ \hline
			0                  &  $ -\frac{64}{245}$   &  $-\frac{256}{175}$  & $ \frac{64}{25}$    & $0$         \\ \hline
			1                  &  $0$   & $ \frac{1536}{1225}$    &  $- \frac{1536}{175}$    &  $\frac{384}{175}$        \\ \hline
			2                  &  $ \frac{64}{245}$    & $ -\frac{64}{245}$    &  $ \frac{64}{5}$    &  $ -\frac{192}{35}$    \\ \hline
			3                  &  $0$  &  $\frac{192}{175}$   &   $ -\frac{2304}{175}$  &  $ \frac{576}{175}$      \\ \hline
			4                  &  $-\frac{1152}{2695}$  &  $-\frac{4608}{1225}$   &   $\frac{1152}{175}$  &  $ 0$      \\ \hline
			5                  &  $0$  &  $\frac{768}{245}$   &   $ 0$  &  $0$      \\ \hline
			6                  &  $\frac{1152}{2695}$  &  $0$   &   $0$  &  $0$      \\ \hline
		\end{tabular}
	\end{subtable}
	\begin{subtable}[h]{0.49 \textwidth}\def\arraystretch{1.25}
		\subcaption{Coefficients $ \mathcal{D}_{m,\ell}^{n}$ associated to $ \,_{2,m}\overline{\chi}_\ell [\dMbar_{a \langle i}^{(n)},\dMbar_{j \rangle a}^{(7-n)}] $}
		\begin{tabular}{|c|c|c|c|}
			\hline
			$\ell$             & \multicolumn{3}{c|}{2}  \\ \hline
			\diagbox{$m$}{$n$} & 0  & 1  & 2     \\ \hline
			0                   &   $0$   & $\frac{576}{175}$    & $0$      \\ \hline
			1                  & $-\frac{576}{1225}$   &  $-\frac{1152}{175}$  &  $\frac{1152}{175}$  \\ \hline
			2                  & $0$  &   $\frac{1152}{245}$  &  $-\frac{576}{35}$      \\ \hline
			3                  & $-\frac{192}{175}$   &  $-\frac{1728}{175}$  &  $\frac{1728}{175}$  \\ \hline
			4                  & $0$   &  $\frac{10368}{1225}$  &  $0$  \\ \hline
			5                  & $\frac{384}{245}$   &  $0$  &  $0$  \\ \hline
		\end{tabular}
	\end{subtable}
	\label{table:coeffs}
\end{table}

\newpage

\bibliography{ListeRef_ToM}

\end{document}